\documentclass[12pt]{article}
\usepackage{graphicx}

\begin{document}

\begin{center}
{\Large \bf Event patterns extracted from transverse momentum and
rapidity spectra of $Z$ bosons and quarkonium states produced in
$pp$ and Pb-Pb collisions at LHC}

\vskip1.0cm

Ya-Hui Chen$^{a}$, Fu-Hu Liu$^{a,}${\footnote{E-mail:
fuhuliu@163.com; fuhuliu@sxu.edu.cn}}, and Roy A.
Lacey$^{b,}${\footnote{E-mail: Roy.Lacey@Stonybrook.edu}}

\vskip0.25cm

{\small\it $^a$Institute of Theoretical Physics, Shanxi
University, Taiyuan, Shanxi 030006, China

$^b$Departments of Chemistry \& Physics, Stony Brook University,
Stony Brook, NY 11794, USA}
\end{center}

\vskip1.0cm

{\bf Abstract:} Transverse momentum ($p_T$) and rapidity ($y$)
spectra of $Z$ bosons and quarkonium states (some charmonium
$c\bar c$ mesons such as $J/\psi$ and $\psi(2S)$, and some
bottomonium $b\bar b$ mesons such as $\Upsilon(1S)$,
$\Upsilon(2S)$, and $\Upsilon(3S)$) produced in proton-proton
($pp$) and lead-lead (Pb-Pb) collisions at the large hadron
collider (LHC) are uniformly described by a hybrid model of
two-component Erlang distribution for $p_T$ spectrum and
two-component Gaussian distribution for $y$ spectrum. The former
distribution results from a multisource thermal model, and the
latter one results from the revised Landau hydrodynamic model. The
modelling results are in agreement with the experimental data
measured in $pp$ collisions at center-of-mass energies
$\sqrt{s}=2.76$ and 7 TeV, and in Pb-Pb collisions at
center-of-mass energy per nucleon pair $\sqrt{s_{NN}}=2.76$ TeV.
Based on the parameter values extracted from $p_T$ and $y$
spectra, the event patterns (particle scatter plots) in
two-dimensional $p_T$-$y$ space and in three-dimensional velocity
space are obtained.
\\

\\


\vskip1.0cm

{\section{Introduction}}

High energy nucleus-nucleus collisions at the relativistic heavy
ion collider (RHIC) [1--4] and large hadron collider (LHC) [5--8]
provide excellent environment and condition of high temperature
and density [9], where a new state of matter, namely the
quark-gluon plasma (QGP) [10--12], is expected to form and to live
for a very short time. It is regretful that the QGP cannot be
directly measured in experiments due to its very short lifetime.
Instead, to understand the formation and properties of QGP, the
distribution laws of final-state particles are studied. Because of
the complexities and difficulties in experiments, the observables
are limited. To obtain more information from limited observables,
we need more modelling and theoretical analyses.

Generally, an interacting system of nucleus-nucleus collisions
undergoes a few stages which include, but are not limited to, the
scattering, color glass condensate, thermalization, hadronization,
chemical equilibrium and freeze-out, kinetic equilibrium and
freeze-out. The distribution laws of final-state particle
transverse momenta and rapidities reflect the situation of
interacting system at the stage of kinetic freeze-out, while the
feed-down corrected yields and the ratios of those yields of
different particles reflect the situation at the stage of chemical
freeze-out. Based on the descriptions of transverse momentum
($p_T$) and rapidity ($y$) spectra, one can extract some
information of transverse excitation and longitudinal expansion of
interacting system. Thus, other information such as the event
pattern (particle scatter plot) in two-dimensional $p_T$-$y$ space
and three-dimensional velocity space can be extracted from
parameters fitted to $p_T$ and $y$ spectra [13, 14].

In peripheral nucleus-nucleus collisions, less nucleons take part
in the interactions. The most peripheral nucleus-nucleus
collisions contain only two nucleons which are from the two
collision nuclei respectively. Proton-proton ($pp$) collisions are
similar to the most peripheral nucleus-nucleus collisions in the
case of neglecting the spectator (cold nuclear) effect. As an
input quantity and a basic collision process, $pp$ collisions can
be used to give comparisons with nucleus-nucleus collisions. We
are interested in both $pp$ collisions and lead-lead (Pb-Pb)
collisions at the LHC.

To understand the stage of kinetic freeze-out in high energy
collisions, we can analyze the $p_T$ and $y$ spectra to obtain the
probability density functions $f_{p_T}(p_T)$ for $p_T$ and
$f_y(y)$ for $y$. Nevertheless, these probability density
functions cannot directly give us a whole and perceptual picture
of the interacting system at the stage of kinetic freeze-out. In
fact, a whole and perceptual picture can help us understand the
interacting mechanisms in detail. Fortunately, we can use the
Monte Carlo method to extract some discrete values of $p_T$ and
$y$ based on $f_{p_T}(p_T)$ and $f_y(y)$. Other quantities such as
energy, momentum components, velocity, and velocity components can
be obtained according to some definitions and assumptions. Because
$f_{p_T}(p_T)$ for $p_T$ and $f_y(y)$ for $y$ are based on
descriptions of experimental spectra, the extracted discrete
values are independent of models.

In this paper, based on a hybrid model of two-component Erlang
distribution for $p_T$ spectrum (which results from a multisource
thermal model [15--17]) and two-component Gaussian distribution
for $y$ spectrum (which results from the Landau hydrodynamic model
and its revisions [18--26]), we analyze together $p_T$ and $y$
spectra of $Z$ bosons and quarkonium states (some charmonium
$c\bar c$ mesons such as $J/\psi$ and $\psi(2S)$, and some
bottomonium $b\bar b$ mesons such as $\Upsilon(1S)$,
$\Upsilon(2S)$, and $\Upsilon(3S)$) produced in $pp$ collisions at
center-of-mass energies $\sqrt{s}=2.76$ and 7 TeV, and in Pb-Pb
collisions at center-of-mass energy per nucleon pair
$\sqrt{s_{NN}}=2.76$ TeV. The modelling results are in agreement
with the experimental data measured at the LHC. Based on the
parameters extracted from $p_T$ and $y$ spectra, the event
patterns (particle scatter plots) at kinetic freeze-out in
two-dimensional $p_T$-$y$ space and in three-dimensional velocity
space are obtained.

The structure of the present work is as followings. The model and
method are shortly described in section 2. Results and discussion
are given in section 3. In section 4, we summarize our main
observations and conclusions.
\\

{\section{The model and method}}

{\it Firstly}, we need modelling descriptions of $p_T$ and $y$
spectra. In the framework of multisource thermal model [15--17],
we can obtain an Erlang distribution or a two-, three- or
multi-component Erlang distribution to fit $p_T$ spectrum.
According to the model, many ($m$) emission sources which stay at
the same excitation state are assumed to form in high energy
collisions. Each (the $i$-th) source is assumed to contribute to
transverse momentum $p_{Ti}$ by an exponential function
\begin{equation}
f_{p_{Ti}}(p_{Ti})=\frac{1}{\langle p_{Ti} \rangle} \exp \bigg(-
\frac{p_{Ti}}{\langle p_{Ti} \rangle} \bigg),
\end{equation}
where $\langle p_{Ti} \rangle$ denotes the average value of
$p_{Ti}$, which results in Eq. (1) to be a probability
distribution and $\int_0^{\infty} f_{p_{Ti}}(p_{Ti}) dp_{Ti}=1$.
The $m$ sources which contribute to $p_T$ resulting in an Erlang
distribution [17]
\begin{equation}
f_{p_T}(p_{T})=\frac{p_T^{m-1}}{(m-1)!\langle p_{Ti} \rangle^m}
\exp \bigg(- \frac{p_{T}}{\langle p_{Ti} \rangle} \bigg)
\end{equation}
which is the folding of $m$ exponential distributions and has the
average transverse momentum $\langle p_T \rangle = m \langle
p_{Ti} \rangle$.

In the case of considering the two-component Erlang distribution,
we have
\begin{equation}
f_{p_T}(p_{T})=\frac{k_1p_T^{m_1-1}}{(m_1-1)!\langle p_{Ti}
\rangle_1^{m_1}} \exp \bigg(- \frac{p_{T}}{\langle p_{Ti}
\rangle_1} \bigg)+\frac{(1-k_1)p_T^{m_2-1}}{(m_2-1)!\langle p_{Ti}
\rangle_2^{m_2}} \exp \bigg(- \frac{p_{T}}{\langle p_{Ti}
\rangle_2} \bigg)
\end{equation}
which has the average transverse momentum $\langle p_T \rangle
=k_1 m_1 \langle p_{Ti} \rangle_1 + (1-k_1)m_2 \langle p_{Ti}
\rangle_2$, where $k_1$ and $1-k_1$ denote the relative
contributions of the first and second components which contribute
in the low- and high-$p_T$ regions respectively, and the
subscripts $1$ and $2$ denote the quantities related to the first
and second components respectively. Eqs. (2) and (3) are
probability distributions which are normalized to 1. When we
compare them with experimental data, normalization constant
($N_{p_T}$) which is used to fit the data is needed.

On $y$ spectrum, we choose the Landau hydrodynamic model and its
revisions [18--26] which are called the revised Landau
hydrodynamic model in the present work. In the model, the
interacting system is described by the hydrodynamics. The $y$
spectrum can be described by a Gaussian function [25, 26]
\begin{equation}
f_y(y)=\frac{1}{\sqrt{2\pi} \sigma_y} \exp \bigg[-
\frac{(y-y_C)^2}{2\sigma_y^2} \bigg],
\end{equation}
where $\sigma_y$ denotes the rapidity distribution width and $y_C$
denotes the mid-rapidity (peak position). In symmetric collisions,
$y_C=0$ is in the center-of-mass reference frame.

In the case of considering the two-component Gaussian function for
$y$ spectrum, we have
\begin{equation}
f_y(y)=\frac{k_B}{\sqrt{2\pi} \sigma_{yB}} \exp \bigg[-
\frac{(y-y_B)^2}{2\sigma_{yB}^2} \bigg] + \frac{1-k_B}{\sqrt{2\pi}
\sigma_{yF}} \exp \bigg[- \frac{(y-y_F)^2}{2\sigma_{yF}^2} \bigg],
\end{equation}
where $k_B$ ($1-k_B$), $y_B$ ($y_F$), and $\sigma_{yB}$
($\sigma_{yF}$) denote respectively the relative contribution,
peak position, and distribution width of the first (second)
component which distributes in the backward (forward) rapidity
region. In symmetric collisions such as $pp$ and Pb-Pb collisions
which are considered in the present work, we have $k_B=1-k_B=0.5$,
$y_B=-y_F$, and $\sigma_{yB}=\sigma_{yF}$. As probability
distributions, Eqs. (4) and (5) are normalized to 1. When we
compare them with experimental data, a normalization constant
($N_y$) which is used to fit the data is needed.

{\it Secondly}, we need discrete values of $p_T$ and $y$. The
related calculation is performed by a Monte Carlo method. Let
$r_i$, $r_{1i}$, $r_{2i}$, and $R_{1,2,\cdots,7}$ denote random
numbers in [0,1]. Eqs. (2)--(5) results in
\begin{equation}
p_T=-\langle p_{Ti} \rangle \sum_{i=1}^m \ln r_i = -\langle p_{Ti}
\rangle \ln \prod_{i=1}^m r_i,
\end{equation}
\begin{equation}
p_T=-\langle p_{Ti} \rangle_1 \sum_{i=1}^{m_1} \ln r_{1i} =
-\langle p_{Ti} \rangle_1 \ln \prod_{i=1}^{m_1} r_{1i}
\end{equation}
for the first component in the low-$p_T$ region, or
\begin{equation}
p_T=-\langle p_{Ti} \rangle_2 \sum_{i=1}^{m_2} \ln r_{2i} =
-\langle p_{Ti} \rangle_2 \ln \prod_{i=1}^{m_2} r_{2i}
\end{equation}
for the second component in the high-$p_T$ region,
\begin{equation}
y=\sigma_y \sqrt{-2\ln R_1} \cos(2\pi R_2) +y_C,
\end{equation}
and
\begin{equation}
y=\sigma_{yB} \sqrt{-2\ln R_3} \cos(2\pi R_4) +y_B
\end{equation}
for the first component in the backward rapidity region, or
\begin{equation}
y=\sigma_{yF} \sqrt{-2\ln R_5} \cos(2\pi R_6) +y_F
\end{equation}
for the second component in the forward rapidity region,
respectively.

It should be clarified that the random numbers used above are
independent in [0,1]. Through the conversions Eqs. (7), (8), (10),
and (11), we can obtain a series of new values which are no longer
independent in [0,1], and obey statistically Eqs. (3) and (5)
respectively. In the Monte Carlo method, Eqs. (6)--(8) are
accustomed expressions which result from the Erlang distribution,
and Eqs. (9)--(11) are accustomed expressions which result from
the Gaussian distribution. If we use $y=\sigma_y \sqrt{-2\ln
R_{1}}$ instead of Eq. (9), we obtain an accustomed expression
which result from the Rayleigh distribution $f(y)=(y/\sigma_y^2)
\exp (-y^2/2\sigma_y^2)$ which is different from the Gaussian
function.

The energy $E$ is given by
\begin{equation}
E=\sqrt{p_T^2+m_0^2}\cosh y,
\end{equation}
where $m_0$ denotes the rest mass of the considered particle. The
$x$-, $y$-, and $z$-components of momentum $p$ are given by
\begin{equation}
p_x=p_T \cos \varphi, \quad p_y=p_T \sin \varphi, \quad
p_z=\sqrt{p_T^2+m_0^2}\sinh y
\end{equation}
respectively, where $\varphi=2\pi R_7$ is the azimuthal angle to
distribute evenly in $[0,2\pi]$. Combining with Eqs. (12) and
(13), we have the velocity
\begin{equation}
\beta=\frac{p}{E}=\frac{\sqrt{p_T^2+p_z^2}}{E} =
\frac{\sqrt{p_x^2+p_y^2+p_z^2}}{E}
\end{equation}
and its components
\begin{equation}
\beta_x=\frac{p_x}{E}, \quad \beta_y=\frac{p_y}{E}, \quad
\beta_z=\frac{p_z}{E}
\end{equation}
respectively. All velocity and its components are in the units of
$c$ which is the speed of light in vacuum and equals to 1 in
natural units.

{\it Thirdly}, we describe the fitting and structuring method step
by step. i) We fit the $p_T$ and $y$ spectra by using Eqs. (3) and
(5) respectively. In the fit, the method of least square method is
used to determine the values of parameters. The minimum $\chi^2$
per degree freedom ($\chi^2$/dof) corresponds to the best values
of parameters. Appropriate increases or decreases in parameters
determine the uncertainties on parameters, where an appropriate
large $\chi^2$/dof is used as a limitation. Because there are
correlations among parameters accounted for, we have to adjust the
parameters and their uncertainties again and again. The best way
is to use a multi-circulation in the calculation by the computer.
ii) Using the best values of parameters, the discrete values of
$p_T$ and $y$ are obtained by Eqs. (7) and (8), as well as (10)
and (11), respectively. The discrete values of momentum and
velocity components are obtained by Eqs. (13) and (15),
respectively. iii) Repeating step ii) many times, we can obtain a
series of discrete values. Then, particle scatter plots, i.e.
event patterns, can be structured by graphic softwares.
\\

{\section{Results and discussion}}

Figure 1 presents (a)(b) transverse momentum spectra,
$d^2\sigma/(dydp_T)$, and (c)(d) rapidity spectra, $d\sigma/dy$,
of $Z$ bosons produced in $pp$ collisions at $\sqrt{s}=2.76$ TeV
for the (a)(c) dimuon ($\mu\mu$) and (b)(d) dielectron ($ee$)
decay channels in $|y|<2.0$ and $|y|<1.44$ respectively, where
$\sigma$ on the vertical axis denotes the cross-section, and the
integral luminosity $L_{\rm int}=5.4$ pb$^{-1}$. The closed
squares represent the experimental data of the CMS Collaboration
[27], and the error bars are only the statistical uncertainties.
The curves are our results calculated by using the two-component
Erlang distribution for $p_T$ spectrum and the two-component
Gaussian distribution for $y$ spectrum, which are the results of
the multisource thermal model [15--17] and the revised Landau
hydrodynamic model [18--26], respectively. The values of free
parameters [$m_1, \langle p_{Ti} \rangle_1, k_1, m_2, \langle
p_{Ti} \rangle_2$, $y_F$ ($=-y_B$) and $\sigma_{yF}$
($=\sigma_{yB}$)], normalization constants ($N_{p_T}$ and $N_y$),
and $\chi^2$/dof are listed in Tables 1 and 2, where the
normalization constant $N_{p_T}$ (or $N_y$) is used to give
comparison between the normalized curve with experimental $p_T$
(or $y$) spectrum. One can see that the results calculated by
using the hybrid model are in agreement with the experimental data
of $Z$ bosons produced in $pp$ collisions at $\sqrt{s}=2.76$ TeV
measured by the CMS Collaboration. In some cases, the values of
$\chi^2$ are very large due to very small experimental errors.

Figure 2 is similar to Figure 1, but it shows the results in Pb-Pb
collisions with 0--100\% centrality at $\sqrt{s_{NN}}=2.76$ TeV,
where the per-event yield $N$ on the vertical axis is used instead
of the cross-section $\sigma$, and $L_{\rm int}=150$
$\mu$b$^{-1}$. The values of free parameters, normalization
constants, and $\chi^2$/dof are listed in Tables 1 and 2. One can
see that the results calculated by using the hybrid model are in
agreement with the experimental data of $Z$ bosons produced in
Pb-Pb collisions at $\sqrt{s_{NN}}=2.76$ TeV measured by the CMS
Collaboration. It should be noted that the ATLAS collaboration has
also published similar experimental results [28, 29]. We can do
similar fits for the spectra of $p_T$ and $y$, which are not
presented in the present work by design due to the similarity.
\\
\\

\begin{figure}
\hskip-1.0cm \begin{center}
\includegraphics[width=10.0cm]{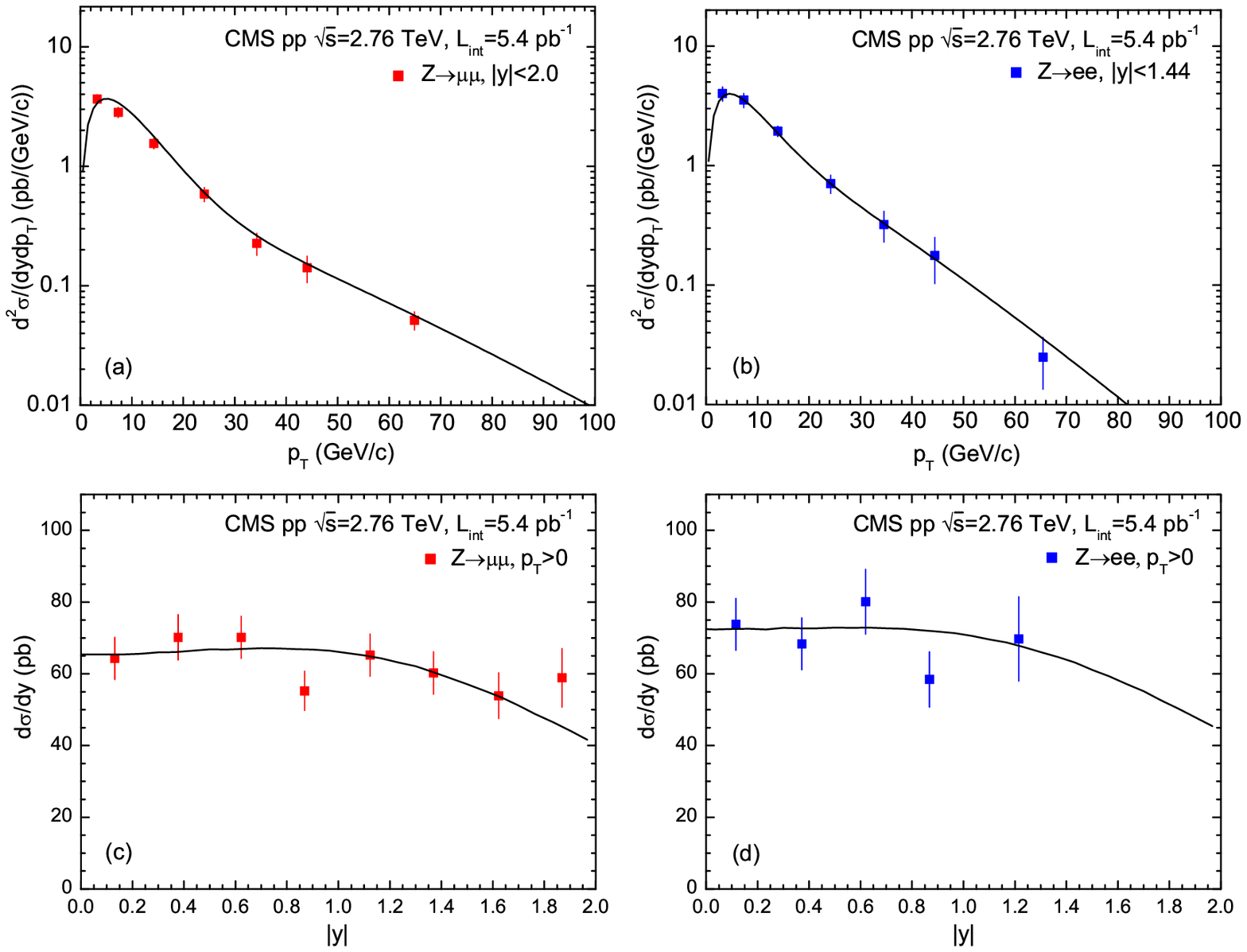}
\end{center}
\vskip-.20cm Fig. 1. (a)(b) Transverse momentum spectra and (c)(d)
rapidity spectra of $Z$ bosons produced in $pp$ collisions at
$\sqrt{s}=2.76$ TeV, for the (a)(c) dimuon ($\mu\mu$) and (b)(d)
dielectron ($ee$) decay channels in $|y|<2.0$ and $|y|<1.44$
respectively, where $\sigma$ on the vertical axis denotes the
cross-section, and the integral luminosity $L_{\rm int}=5.4$
pb$^{-1}$. The closed squares represent the experimental data of
the CMS Collaboration [27], and the error bars are only the
statistical uncertainties. The curves are our results calculated
by using the (a)(b) two-component Erlang distribution and (c)(d)
two-component Gaussian distribution, respectively.
\end{figure}

\begin{figure}
\hskip-1.0cm \begin{center}
\includegraphics[width=10.0cm]{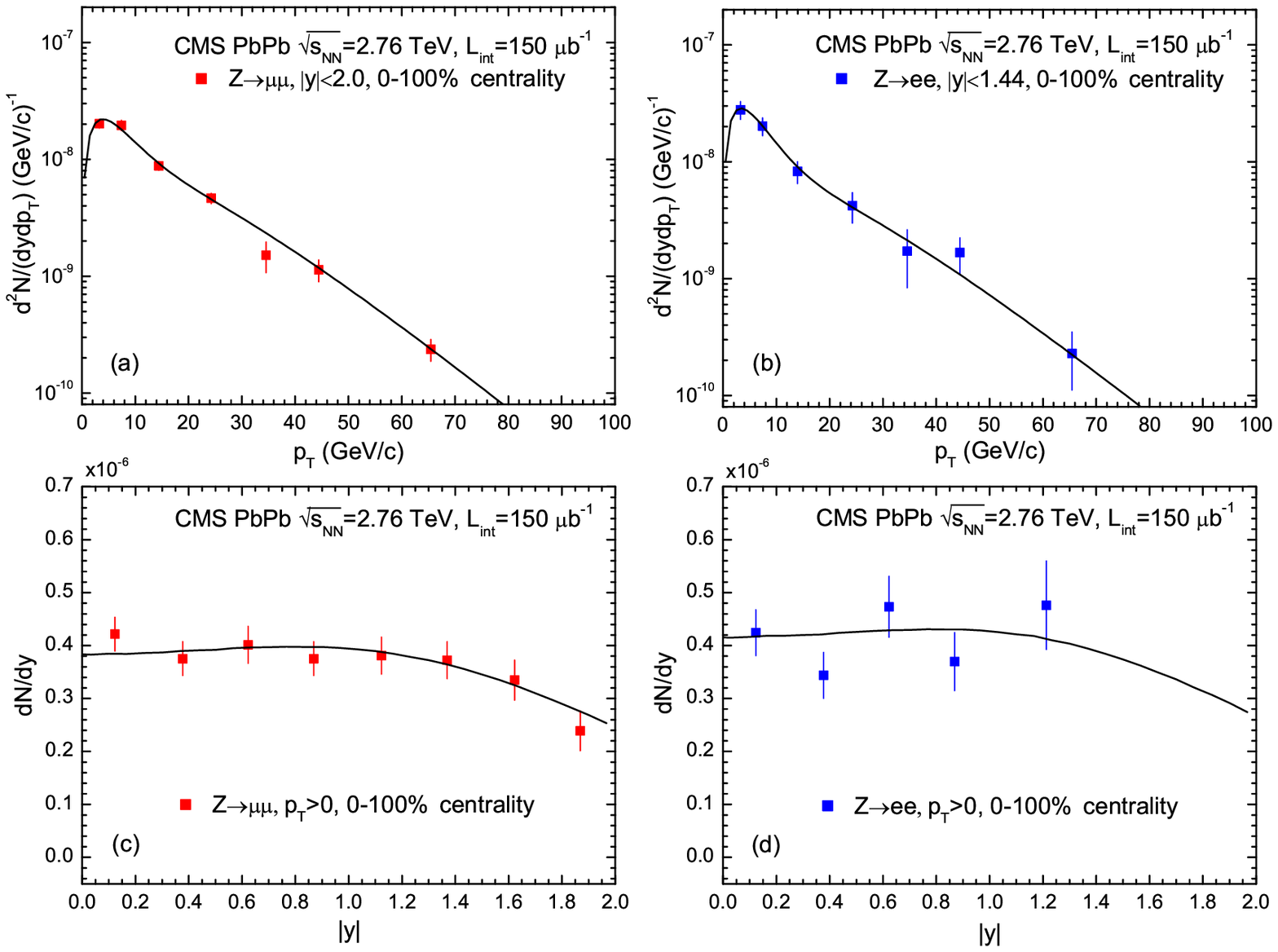}
\end{center}
\vskip-.20cm Fig. 2.  Same as Figure 1, but showing the results in
Pb-Pb collisions with 0--100\% centrality at $\sqrt{s_{NN}}=2.76$
TeV, where the per-event yield $N$ on the vertical axis is used
instead of the cross-section $\sigma$, and $L_{\rm int}=150$
$\mu$b$^{-1}$.
\end{figure}

{\tiny {Table 1. Values of free parameters ($m_{1}$, $\langle
p_{Ti} \rangle_1$, $k_{1}$, $m_{2}$, and $\langle p_{Ti}
\rangle_2$), normalization constant ($N_{p_T}$), and $\chi^2$/dof
corresponding to the two-component Erlang distribution in Figures
1(a), 1(b), 2(a), 2(b), 5(a), 5(b), 8(a), 11(a) and 11(c), where
both the uncertainties on $m_1$ and $m_2$ are 0 which are not
listed in the columns. In some cases, dof is less than 1.
{%
\begin{center}
\begin{tabular}{ccccccccc}
\hline\hline Figure & Type & $m_{1}$ & $\langle p_{Ti} \rangle_1$ (GeV/$c$) & $k_{1}$  & $m_{2}$ & $\langle p_{Ti} \rangle_2$ (GeV/$c$) & $N_{p_T}$ & $\chi^2$/dof \\
\hline
Figure 1(a) & $pp$, $Z\rightarrow\mu\mu$  & 2 & $4.85\pm0.20$ & $0.78\pm0.02$ & 2 & $15.80\pm0.79$& $55.00\pm2.75$ & 22.945/1 \\
Figure 1(b) & $pp$, $Z\rightarrow ee$     & 2 & $4.10\pm0.20$ & $0.60\pm0.02$ & 2 & $11.00\pm0.55$& $63.00\pm3.15$ & 7.495/1 \\
\hline
Figure 2(a) & Pb-Pb, $Z\rightarrow\mu\mu$ & 2 & $3.35\pm0.17$ & $0.45\pm0.02$ & 2 & $10.60\pm0.53$& $(3.57\pm0.18)\times10^{-7}$ & 10.850/1 \\
Figure 2(b) & Pb-Pb, $Z\rightarrow ee$    & 2 & $3.05\pm0.15$ & $0.54\pm0.02$ & 2 & $10.70\pm0.53$& $(3.85\pm0.19)\times10^{-7}$ & 2.606/1 \\
\hline
Figure 5(a) & $pp$, $J/\psi$              & 2 & $1.15\pm0.06$ & $0.95\pm0.02$ & 2 & $2.34\pm0.12$ & $5.00\pm0.25$  & 41.272/28 \\
Figure 5(b) & $pp$, $\psi(2S)$            & 2 & $1.30\pm0.07$ & $0.90\pm0.02$ & 2 & $2.50\pm0.13$ & $0.75\pm0.04$  & 2.587/15 \\
\hline
Figure 8(a) & $pp$, $\Upsilon(1S)$        & 2 & $2.30\pm0.12$ & $0.65\pm0.02$ & 3 & $1.75\pm0.09$ & $2.38\pm0.12$  & 1.997/9 \\
            & $pp$, $\Upsilon(2S)$        & 3 & $1.78\pm0.09$ & $0.68\pm0.02$ & 2 & $2.80\pm0.14$ & $0.56\pm0.03$  & 0.506/6 \\
            & $pp$, $\Upsilon(3S)$        & 3 & $2.00\pm0.10$ & $0.40\pm0.02$ & 2 & $3.20\pm0.16$ & $0.29\pm0.01$  & 5.800/5 \\
\hline
Figure 11(a)& Pb-Pb, $J/\psi$, 0--20\%    & 2 & $0.89\pm0.04$ & $0.50\pm0.02$ & 3 & $0.69\pm0.03$ & $(2.50\pm0.13)\times10^{-2}$ & 12.872/7 \\
            & Pb-Pb, $J/\psi$, 20--40\%   & 2 & $0.90\pm0.05$ & $0.40\pm0.02$ & 3 & $0.78\pm0.04$ & $(9.00\pm0.45)\times10^{-3}$ & 18.291/7 \\
            & Pb-Pb, $J/\psi$, 40--90\%   & 2 & $0.93\pm0.05$ & $0.40\pm0.02$ & 3 & $0.78\pm0.04$ & $(1.60\pm0.08)\times10^{-3}$ & 29.495/7 \\
Figure 11(c)& Pb-Pb, $\Upsilon(1S)$       & 3 & $1.50\pm0.08$ & $0.35\pm0.02$ & 2 & $3.00\pm0.15$ & $0.30\pm0.02$                & 11.024/$-1$ \\
            & Pb-Pb, $\Upsilon(2S)$       & 3 & $1.50\pm0.08$ & $0.25\pm0.02$ & 2 & $4.00\pm0.20$ & $(1.50\pm0.08)\times10^{-2}$ & 0.287/$-3$ \\
\hline
\end{tabular}
\end{center}
}} }

\vskip1.5cm
{\scriptsize {Table 2. Values of free parameter [$y_F$ ($=-y_B$)
and $\sigma_{yF}$ ($=\sigma_{yB}$)], normalization constant
($N_y$), and $\chi^2$/dof corresponding to the two-component
Gaussian distribution in Figures 1(c), 1(d), 2(c), 2(d), 5(c),
5(d), 8(b), 8(c), 8(d), 11(b), and 11(d). In some cases, dof is
less than 1.
{%
\begin{center}
\begin{tabular}{ccccccc}
\hline\hline  Figure & Type & $y_F$ ($=-y_B$) & $\sigma_{yF}$ ($=\sigma_{yB}$) & $N_{y}$ & $\chi^2$/dof \\
\hline
Figure 1(c) &$pp$, $Z\rightarrow\mu\mu$ & $1.10\pm0.03$ & $1.00\pm0.05$ & $300\pm15$     & 8.622/5 \\
Figure 1(d) &$pp$, $Z\rightarrow ee$    & $1.10\pm0.03$ & $1.05\pm0.05$ & $330\pm17$     & 2.074/2 \\
\hline
Figure 2(c) &Pb-Pb, $Z\rightarrow\mu\mu$& $1.12\pm0.03$ & $1.00\pm0.05$ & $1.80\pm0.09$  & 2.860/5 \\
Figure 2(d) &Pb-Pb, $Z\rightarrow ee$   & $1.12\pm0.03$ & $1.00\pm0.05$ & $1.95\pm0.10$  & 5.626/2 \\
\hline
Figure 5(c) &$pp$, $J/\psi$             & $1.90\pm0.03$ & $1.82\pm0.09$ & $54\pm3$       & 8.141/14 \\
Figure 5(d) &$pp$, $\psi(2S)$           & $1.90\pm0.03$ & $1.85\pm0.10$ & $8.8\pm0.4$    & 1.338/3 \\
\hline
Figure 8(b) &$pp$, $\Upsilon(1S)$       & $1.58\pm0.03$ & $1.65\pm0.08$ & $13.00\pm0.65$ & 0.492/2 \\
            &$pp$, $\Upsilon(2S)$       & $1.55\pm0.03$ & $1.48\pm0.07$ & $3.20\pm0.16$  & 0.657/2 \\
            &$pp$, $\Upsilon(3S)$       & $1.55\pm0.03$ & $1.60\pm0.08$ & $1.50\pm0.08$  & 0.395/2 \\
Figure 8(c) &$pp$, $\Upsilon(1S)$       & $1.58\pm0.03$ & $1.65\pm0.08$ & $510\pm26$     & 22.156/17 \\
Figure 8(d) &$pp$, $\Upsilon(2S)$       & $1.55\pm0.03$ & $1.48\pm0.07$ & $160\pm8$      & 5.760/9 \\
\hline
Figure 11(b)&Pb-Pb, $J/\psi$, inclusive & $1.40\pm0.03$ & $1.70\pm0.09$ & $17.00\pm0.85$ & 0.504/0 \\
            &Pb-Pb, $J/\psi$, prompt    & $1.51\pm0.03$ & $1.70\pm0.09$ & $13.70\pm0.69$ & 0.899/0 \\
Figure 11(d)&Pb-Pb, $\Upsilon(1S)$      & $1.35\pm0.03$ & $1.25\pm0.06$ & $1.95\pm0.10$  & 1.070/3 \\
            &Pb-Pb, $\Upsilon(2S)$      & $1.38\pm0.03$ & $1.22\pm0.06$ & $0.16\pm0.008$& 0.085/$-1$ \\
\hline
\end{tabular}
\end{center}
}} }

\vskip0.5cm

Based on the parameter values obtained from Figures 1 and 2, we
can perform a Monte Carlo calculation and obtain a series of
kinematic quantities. For example, event patterns in
two-dimensional $p_T-y$ space are presented in Figure 3 for (a)(c)
$Z\rightarrow\mu\mu$ and (b)(d) $Z\rightarrow ee$ channels in
(a)(b) $pp$ collisions at $\sqrt{s}=2.76$ TeV and (c)(d) Pb-Pb
collisions at $\sqrt{s_{NN}}=2.76$ TeV. The total number of
particles for each panel is 1000. The squares and circles
represent the contributions of the first and second components in
the two-component Erlang distribution for $p_T$ spectrum
respectively. Most particles appear in the region of low-$p_T$ and
low-$|y|$. The values of root-mean-squares
($\sqrt{\overline{p_T^2}}$ for $p_T$ and $\sqrt{\overline{y^2}}$
for $y$) are listed in Table 3. In the calculation, the
contributions of the first and second components in the
two-component Gaussian distribution for $y$ spectrum are not
distinguished by design.

Figure 4 is another example of event pattern in three-dimensional
velocity ($\beta_x-\beta_y-\beta_z$) space for (a)(c)
$Z\rightarrow\mu\mu$ and (b)(d) $Z\rightarrow ee$ channels in
(a)(b) $pp$ collisions at $\sqrt{s}=2.76$ TeV and (c)(d) Pb-Pb
collisions at $\sqrt{s_{NN}}=2.76$ TeV. The total number of
particles for each panel is 1000. The red and blue globules
represent the contributions of the first and second components in
the two-component Erlang distribution for $p_T$ spectrum
respectively. Most particles appear in the region of
low-$|\beta_x|$ and low-$|\beta_y|$. The values of
root-mean-squares ($\sqrt{\overline{\beta_x^2}}$ for $\beta_x$,
$\sqrt{\overline{\beta_y^2}}$ for $\beta_y$, and
$\sqrt{\overline{\beta_z^2}}$ for $\beta_z$) and the maximum
$|\beta_x|$, $|\beta_y|$, and $|\beta_z|$ ($|\beta_x|_{\max}$,
$|\beta_y|_{\max}$, and $|\beta_z|_{\max}$) are listed in Table 4.
One can see that $\sqrt{\overline{\beta_x^2}} \approx
\sqrt{\overline{\beta_y^2}} \ll \sqrt{\overline{\beta_z^2}}$, and
$|\beta_x|_{\max} \approx |\beta_y|_{\max}<|\beta_z|_{\max}$. The
event pattern in velocity (or coordinate) space looks like a rough
cylinder along $oz$ axis (the beam direction) when it emit $Z$
bosons, and the maximum velocity surface (the surface consisted of
the maximum velocities in different directions) is a fat cylinder
which has the length being 1.6--2.2 times of diameter. As for
Figure 3, the contributions of the first and second components in
the two-component Gaussian distribution for $y$ spectrum are not
distinguished by design.

In Figure 5, the spectra of inclusive (a) $J/\psi$ $p_T$, (b)
$\psi(2S)$ $p_T$, (c) $J/\psi$ $y$, and (d) $\psi(2S)$ $y$ mesons
in $pp$ collisions at $\sqrt{s}=7$ TeV are displayed. Different
solid symbols represent different data measured by the ALICE
[30--32] and LHCb collaborations [33, 34] in different conditions
shown in the panels, and the error bars are only the statistical
uncertainties. The curves for $p_T$ and $y$ spectra are our
results calculated by using the two-component Erlang distribution
for $p_T$ spectrum and the two-component Gaussian distribution for
$y$ spectrum respectively. The values of free parameters,
normalization constants, and $\chi^2$/dof are listed in Tables 1
and 2. One can see that the results calculated by using the hybrid
model are in agreement with the experimental data of quarkonium
states (charmonium $c\bar c$ mesons $J/\psi$ and $\psi(2S)$)
produced in $pp$ collisions at $\sqrt{s}=7$ TeV measured by the
ALICE and LHCb Collaborations.

Based on the parameter values obtained from Figure 5, we can
obtain some event patterns for $J/\psi$ and $\psi(2S)$. In Figures
6 and 7, the event patterns in $p_T-y$ and
$\beta_x-\beta_y-\beta_z$ spaces for (a) inclusive $J/\psi$ mesons
and (b) inclusive $\psi(2S)$ mesons are given. The red squares and
globules represent the contribution of the first component, and
the blue circles and globules represent the contribution of the
second component, in the two-component Erlang distribution for
$p_T$ spectrum. The corresponding values of root-mean squares and
the maximum velocity components are listed in Tables 3 and 4. One
can see that $\sqrt{\overline{p_T^2}}$ ($\sqrt{\overline{y^2}}$)
for charmonium $c\bar c$ mesons is less (greater) than that for
$Z$ bosons. Once again, $\sqrt{\overline{\beta_x^2}} \approx
\sqrt{\overline{\beta_y^2}} \ll \sqrt{\overline{\beta_z^2}}$, and
$|\beta_x|_{\max} \approx |\beta_y|_{\max}<|\beta_z|_{\max}$. At
the same time, $\sqrt{\overline{\beta_x^2}}$ and
$|\beta_x|_{\max}$ ($\sqrt{\overline{\beta_y^2}}$ and
$|\beta_y|_{\max}$, $\sqrt{\overline{\beta_z^2}}$) for charmonium
$c\bar c$ mesons are larger than those for $Z$ bosons, and the two
types of particles have nearly the same $|\beta_z|_{\max}$. The
event pattern in velocity (or coordinate) space looks like a rough
cylinder along the beam direction when it emit charmonium $c\bar
c$ mesons, and the maximum velocity surface is a fat cylinder
which has the length being 1.2--1.4 times of diameter.

\begin{figure}
\hskip-1.0cm \begin{center}
\includegraphics[width=10.0cm]{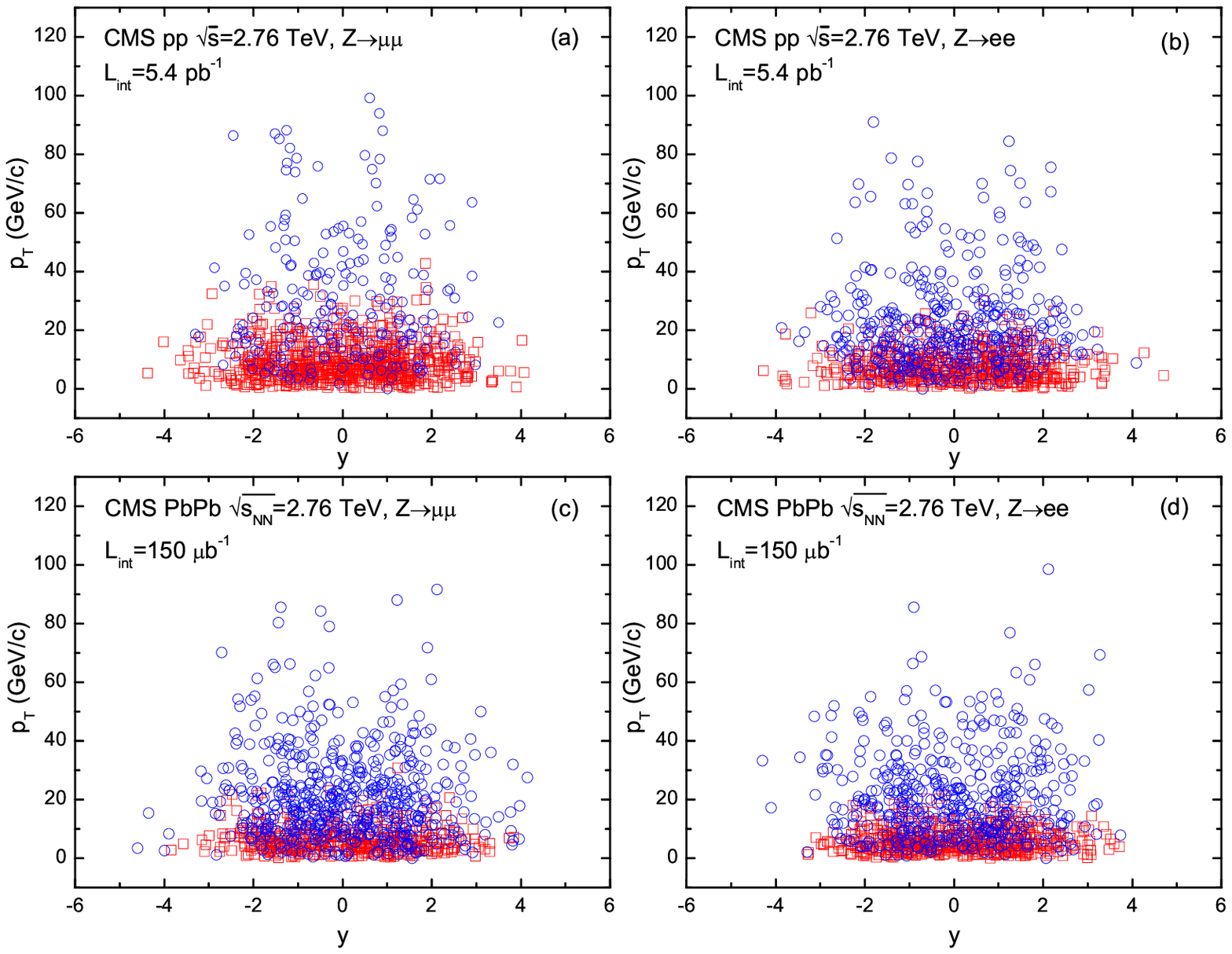}
\end{center}
\vskip-.20cm {\small Fig. 3. Event patterns (particle scatter
plots) in two-dimensional $p_T-y$ space in (a)(b) $pp$ collisions
at $\sqrt{s}=2.76$ TeV and (c)(d) Pb-Pb collisions at
$\sqrt{s_{NN}}=2.76$ TeV, for the (a)(c) dimuon and (b)(d)
dielectron decay channels. The number of $Z$ bosons for each panel
is 1000. The squares and circles represent the results of the
first and second components in the two-component Erlang
distribution for $p_T$, respectively. The contributions of the
first and second components in the two-component Gaussian
distribution for $y$ are not distinguished by design.}
\end{figure}

\vskip1.0cm

{\scriptsize {Table 3. Values of the root-mean-squares
$\sqrt{\overline{p_T^2}}$ for $p_T$ and $\sqrt{\overline{y^2}}$
for $y$ corresponding to the particle scatter plots in Figures 3,
6, 9, and 12.
{%
\begin{center}
\begin{tabular}{cccc}
\hline \hline Figure & Type & $\sqrt{\overline{p_T^2}}$ (GeV/$c$) & $\sqrt{\overline{y^2}}$  \\
\hline
Figure 3(a)  &$pp$, $Z\rightarrow\mu\mu$ & $18.91\pm0.72$ & $1.50\pm0.03$ \\
Figure 3(b)  &$pp$, $Z\rightarrow ee$    & $19.05\pm0.69$ & $1.49\pm0.03$ \\
Figure 3(c)  &Pb-Pb, $Z\rightarrow\mu\mu$& $20.38\pm0.66$ & $1.50\pm0.03$ \\
Figure 3(d)  &Pb-Pb, $Z\rightarrow ee$   & $18.82\pm0.64$ & $1.47\pm0.03$ \\
\hline
Figure 6(a)  &$pp$, $J/\psi$             & $3.11\pm0.09$  & $2.68\pm0.05$ \\
Figure 6(b)  &$pp$, $\psi(2S)$           & $3.40\pm0.09$  & $2.67\pm0.05$ \\
\hline
Figure 9(a)  &$pp$, $\Upsilon(1S)$       & $5.53\pm0.11$  & $2.26\pm0.05$ \\
Figure 9(b)  &$pp$, $\Upsilon(2S)$       & $6.03\pm0.11$  & $2.13\pm0.04$ \\
Figure 9(c)  &$pp$, $\Upsilon(3S)$       & $6.49\pm0.11$  & $2.23\pm0.04$ \\
\hline
Figure 12(a) &Pb-Pb, $\Upsilon(1S)$      & $6.26\pm0.14$  & $1.88\pm0.04$ \\
Figure 12(b) &Pb-Pb, $\Upsilon(2S)$      & $7.80\pm0.15$  & $1.83\pm0.03$ \\
\hline
\end{tabular}%
\end{center}
}} }

\vskip0.5cm

In the upper panel of Figure 8, the spectra of (a) $p_T$ for
$\Upsilon(1S)$, $\Upsilon(2S)$, and $\Upsilon(3S)$ and (b) $y$ in
backward (forward) region for $\Upsilon(1S)$, $\Upsilon(2S)$, and
$\Upsilon(3S)$ produced in $pp$ collisions at $\sqrt{s}=7$ TeV are
presented, where $\sigma^{iS}$ and $B^{iS}$ on the vertical axis
denote the cross-section and branching fractions of $\Upsilon(iS)$
($i=1$, 2, and 3), respectively. The closed circles, squares, and
triangles represent the experimental data of $\Upsilon(1S)$,
$\Upsilon(2S)$, and $\Upsilon(3S)$, respectively, measured by the
LHCb Collaboration [35], and the error bars are the total
uncertainties which combine the systematic and statistical
uncertainties to be equal to their root-sum-square. The curves for
$p_T$ and $y$ spectra are our results calculated by using the
two-component Erlang distribution and the two-component Gaussian
distribution respectively. The values of free parameters,
normalization constants, and $\chi^2$/dof are listed in Tables 1
and 2. One can see that the results calculated by using the hybrid
model are in agreement with the experimental data of quarkonium
states (bottomonium $b\bar b$ mesons $\Upsilon(1S)$,
$\Upsilon(2S)$, and $\Upsilon(3S)$) measured by the LHCb
Collaboration.

\begin{figure}
\hskip-1.0cm \begin{center}
\includegraphics[width=10.0cm]{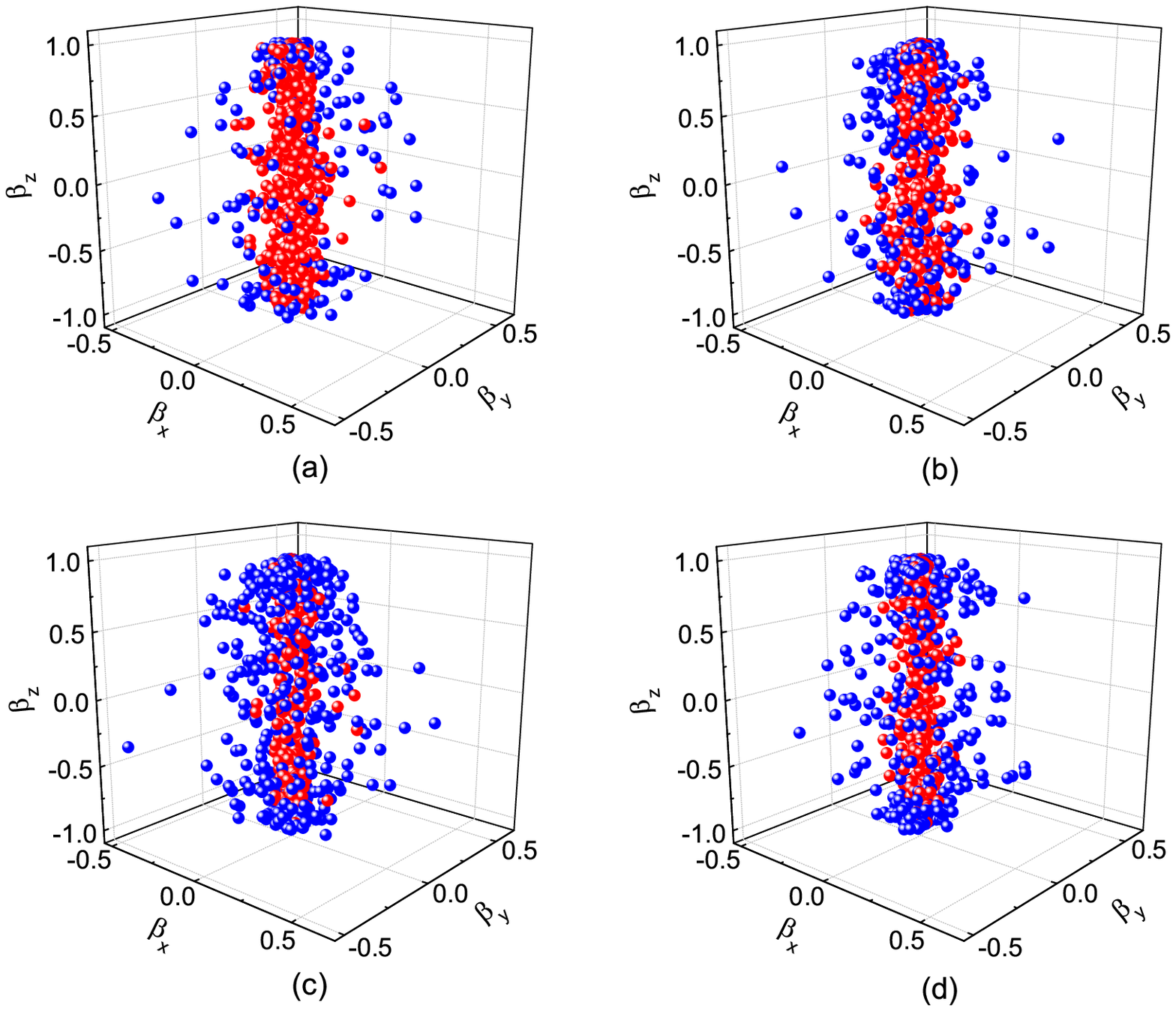}
\end{center}
\vskip-.20cm {\small Fig. 4. Event patterns in three-dimensional
velocity ($\beta_x-\beta_y-\beta_z$) space in (a)(b) $pp$
collisions at $\sqrt{s}=2.76$ TeV and (c)(d) Pb-Pb collisions at
$\sqrt{s_{NN}}=2.76$ TeV, for the (a)(c) dimuon and (b)(d)
dielectron decay channels. The number of $Z$ bosons for each panel
is 1000. The red and blue globules represent the results of the
first and second components in the two-component Erlang
distribution for $p_T$, respectively. The contributions of the
first and second components in the two-component Gaussian
distribution for $y$ are not distinguished by design.}
\end{figure}

\vskip0.5cm
{\scriptsize {Table 4. Values of the root-mean-squares
$\sqrt{\overline{\beta_x^2}}$ for $\beta_x$,
$\sqrt{\overline{\beta_y^2}}$ for $\beta_y$, and
$\sqrt{\overline{\beta_z^2}}$ for $\beta_z$, as well as the
maximum $|\beta_x|$, $|\beta_y|$, and $|\beta_z|$
($|\beta_x|_{\max}$, $|\beta_y|_{\max}$, and $|\beta_z|_{\max}$)
corresponding to the particle scatter plots in Figures 4, 7, 10,
and 13.
{%
\begin{center}
\begin{tabular}{cccccccc}
\hline\hline Figure & Type & $\sqrt{\overline{\beta_x^2}}$ ($c$) & $\sqrt{\overline{\beta_y^2}}$ ($c$) & $\sqrt{\overline{\beta_z^2}}$ ($c$) & $|\beta_x|_{\max}$ ($c$) & $|\beta_y|_{\max}$ ($c$) & $|\beta_z|_{\max}$ ($c$) \\
\hline
Figure 4(a)    &$pp$, $Z\rightarrow\mu\mu$ & $0.08\pm0.01$ & $0.09\pm0.01$ & $0.76\pm0.01$ & 0.56 & 0.47 & 1.00  \\
Figure 4(b)    &$pp$, $Z\rightarrow ee$    & $0.09\pm0.01$ & $0.08\pm0.01$ & $0.76\pm0.01$ & 0.56 & 0.60 & 1.00  \\
Figure 4(c)    &Pb-Pb, $Z\rightarrow\mu\mu$& $0.09\pm0.01$ & $0.10\pm0.01$ & $0.75\pm0.01$ & 0.65 & 0.54 & 1.00  \\
Figure 4(d)    &Pb-Pb, $Z\rightarrow ee$   & $0.08\pm0.01$ & $0.08\pm0.01$ & $0.76\pm0.01$ & 0.54 & 0.45 & 1.00  \\
\hline
Figure 7(a)    &$pp$, $J/\psi$             & $0.16\pm0.01$ & $0.17\pm0.01$ & $0.90\pm0.01$ & 0.83 & 0.76 & 1.00 \\
Figure 7(b)    &$pp$, $\psi(2S)$           & $0.20\pm0.01$ & $0.21\pm0.01$ & $0.88\pm0.01$ & 0.84 & 0.77 & 1.00 \\
\hline
Figure 10(a)   &$pp$, $\Upsilon(1S)$       & $0.14\pm0.01$ & $0.14\pm0.01$ & $0.85\pm0.01$ & 0.79 & 0.61 & 1.00 \\
Figure 10(b)   &$pp$, $\Upsilon(2S)$       & $0.13\pm0.01$ & $0.13\pm0.01$ & $0.83\pm0.01$ & 0.67 & 0.67 & 1.00 \\
Figure 10(c)   &$pp$, $\Upsilon(3S)$       & $0.19\pm0.01$ & $0.19\pm0.01$ & $0.84\pm0.01$ & 0.74 & 0.76 & 1.00 \\
\hline
Figure 13(a)   &Pb-Pb, $\Upsilon(1S)$      & $0.19\pm0.01$ & $0.17\pm0.01$ & $0.81\pm0.01$ & 0.81 & 0.78 & 1.00 \\
Figure 13(b)   &Pb-Pb, $\Upsilon(2S)$      & $0.21\pm0.01$ & $0.22\pm0.01$ & $0.81\pm0.01$ & 0.79 & 0.83 & 1.00 \\
\hline
\end{tabular}
\end{center}
}} }

\vskip0.5cm

\begin{figure}
\hskip-1.0cm \begin{center}
\includegraphics[width=10.0cm]{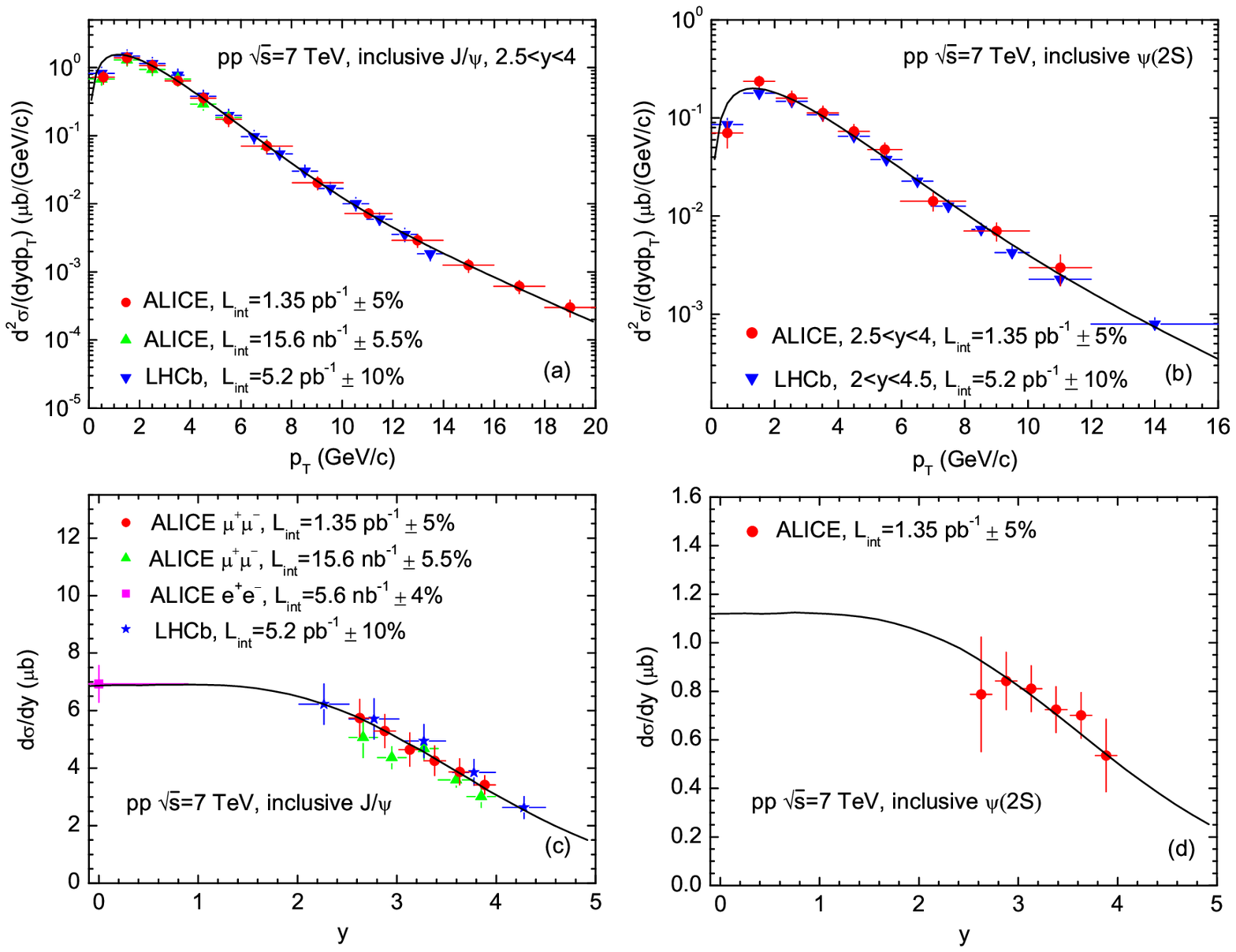}
\end{center}
\vskip-.20cm Fig. 5. (a)(b) Transverse momentum spectra and (c)(d)
rapidity spectra of (a)(c) inclusive $J/\psi$ mesons and (b)(d)
inclusive $\psi(2S)$ mesons produced in $pp$ collisions at
$\sqrt{s}=7$ TeV. The closed symbols represent the experimental
data of the ALICE [30--32] and LHCb Collaboration [33, 34] in
different conditions shown in the panels, and the error bars are
only the statistical uncertainties. The curves are our results
calculated by using the hybrid model.
\end{figure}

To study further the $y$ spectra of bottomonium $b\bar b$ mesons,
Figures 8(c) and 8(d) give the results for $\Upsilon(1S)$ and
$\Upsilon(2S)$ in $pp$ collisions at 7 TeV respectively. Different
closed symbols represent different data measured by the ALICE
[30], LHCb [35], and CMS collaborations [36, 37] in different
conditions shown in the panels, and the error bars are only the
systematic uncertainties. The curves are our results calculated by
using the two-component Gaussian distribution. The values of free
parameters, normalization constants, and $\chi^2$/dof are listed
in Table 2. One can see that the results calculated by using the
revised Landau hydrodynamic model are in indeed agreement with the
experimental data of bottomonium $b\bar b$ mesons $\Upsilon(1S)$
and $\Upsilon(2S)$.

\begin{figure}
\hskip-1.0cm \begin{center}
\includegraphics[width=10.0cm]{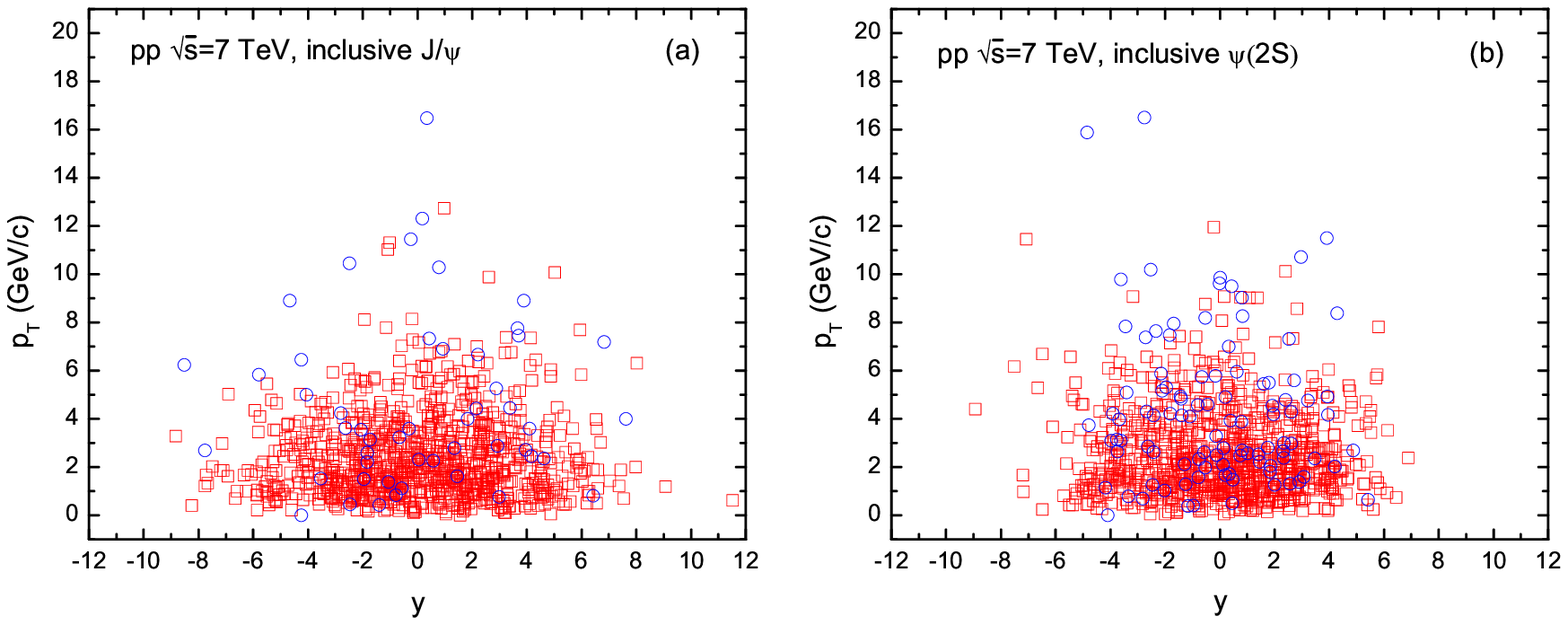}
\end{center}
\vskip-.20cm Fig. 6. Same as Figure 3, but showing the results for
(a) inclusive $J/\psi$ mesons and (b) inclusive $\psi(2S)$ mesons
produced in $pp$ collisions at $\sqrt{s}=7$ TeV.
\end{figure}

Based on the parameter values obtained from Figures 8(a) and 8(b),
we can obtain some event patterns for $\Upsilon(1S)$,
$\Upsilon(2S)$, and $\Upsilon(3S)$. In Figures 9 and 10, the event
patterns in $p_T-y$ and $\beta_x-\beta_y-\beta_z$ spaces for (a)
$\Upsilon(1S)$, (b) $\Upsilon(2S)$, and (c) $\Upsilon(3S)$ are
given. The red squares and globules represent the contribution of
the first component, and the blue circles and globules represent
the contribution of the second component, in the two-component
Erlang distribution. The corresponding values of root-mean squares
and the maximum velocity components are listed in Tables 3 and 4.
One can see that $\sqrt{\overline{p_T^2}}$
($\sqrt{\overline{y^2}}$) for bottomonium $b\bar b$ mesons is
close to that for charmonium $c\bar c$ mesons, and less (greater)
than that for $Z$ bosons. Once more, $\sqrt{\overline{\beta_x^2}}
\approx \sqrt{\overline{\beta_y^2}} \ll
\sqrt{\overline{\beta_z^2}}$, and $|\beta_x|_{\max} \approx
|\beta_y|_{\max}<|\beta_z|_{\max}$. At the same time,
$\sqrt{\overline{\beta_x^2}}$ and $|\beta_x|_{\max}$
($\sqrt{\overline{\beta_y^2}}$ and $|\beta_y|_{\max}$,
$\sqrt{\overline{\beta_z^2}}$) for bottomonium $b\bar b$ mesons
are close to those for charmonium $c\bar c$ mesons, and larger
than those for $Z$ bosons. The three types of particles have
nearly the same $|\beta_z|_{\max}$. The event pattern in velocity
(or coordinate) space looks like a rough cylinder along the beam
direction when it emit bottomonium $b\bar b$ mesons, and the
maximum velocity surface is a fat cylinder which has the length
being 1.3--1.6 times of diameter.

\begin{figure}
\hskip-1.0cm \begin{center}
\includegraphics[width=10.0cm]{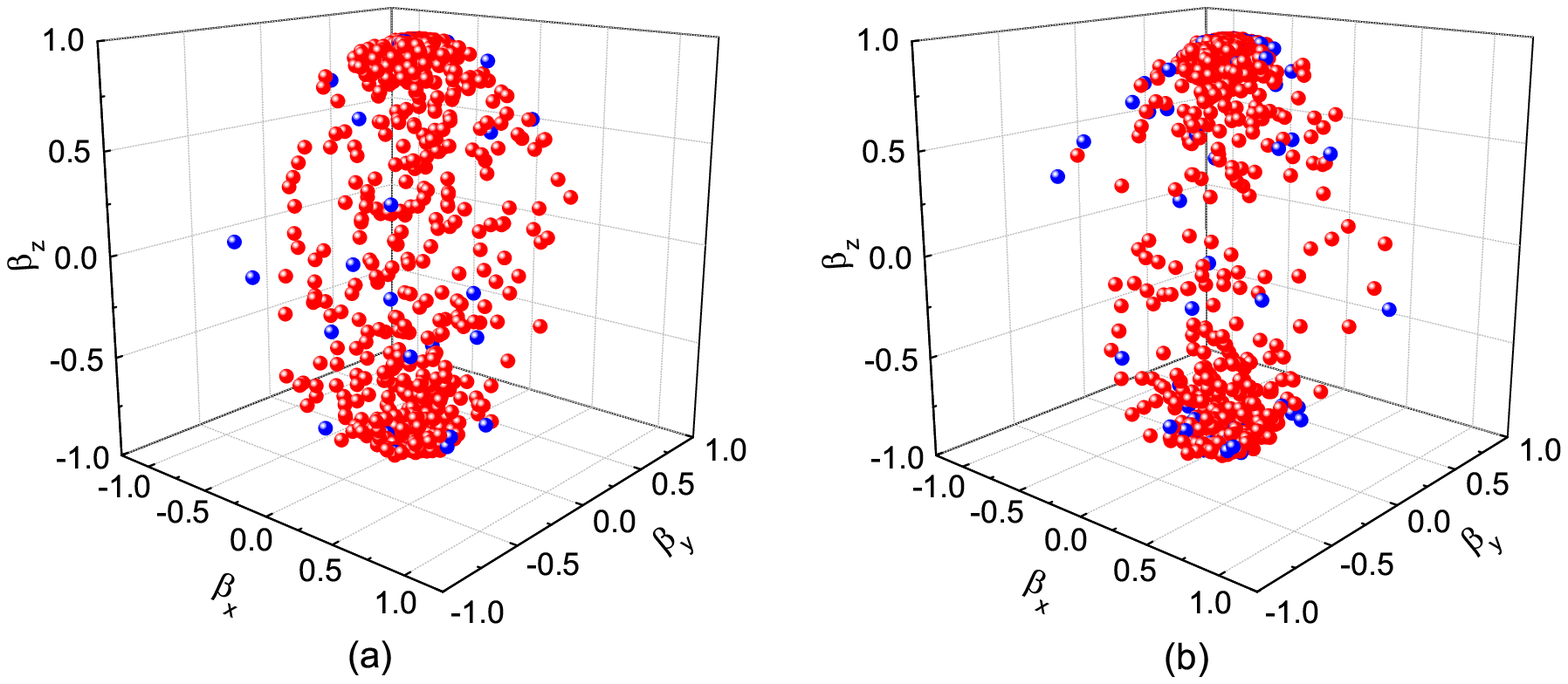}
\end{center}
\vskip-.20cm Fig. 7. Same as Figure 4, but showing the results for
(a) inclusive $J/\psi$ mesons and (b) inclusive $\psi(2S)$ mesons
produced in $pp$ collisions at $\sqrt{s}=7$ TeV.
\end{figure}

\begin{figure}
\hskip-1.0cm \begin{center}
\includegraphics[width=10.0cm]{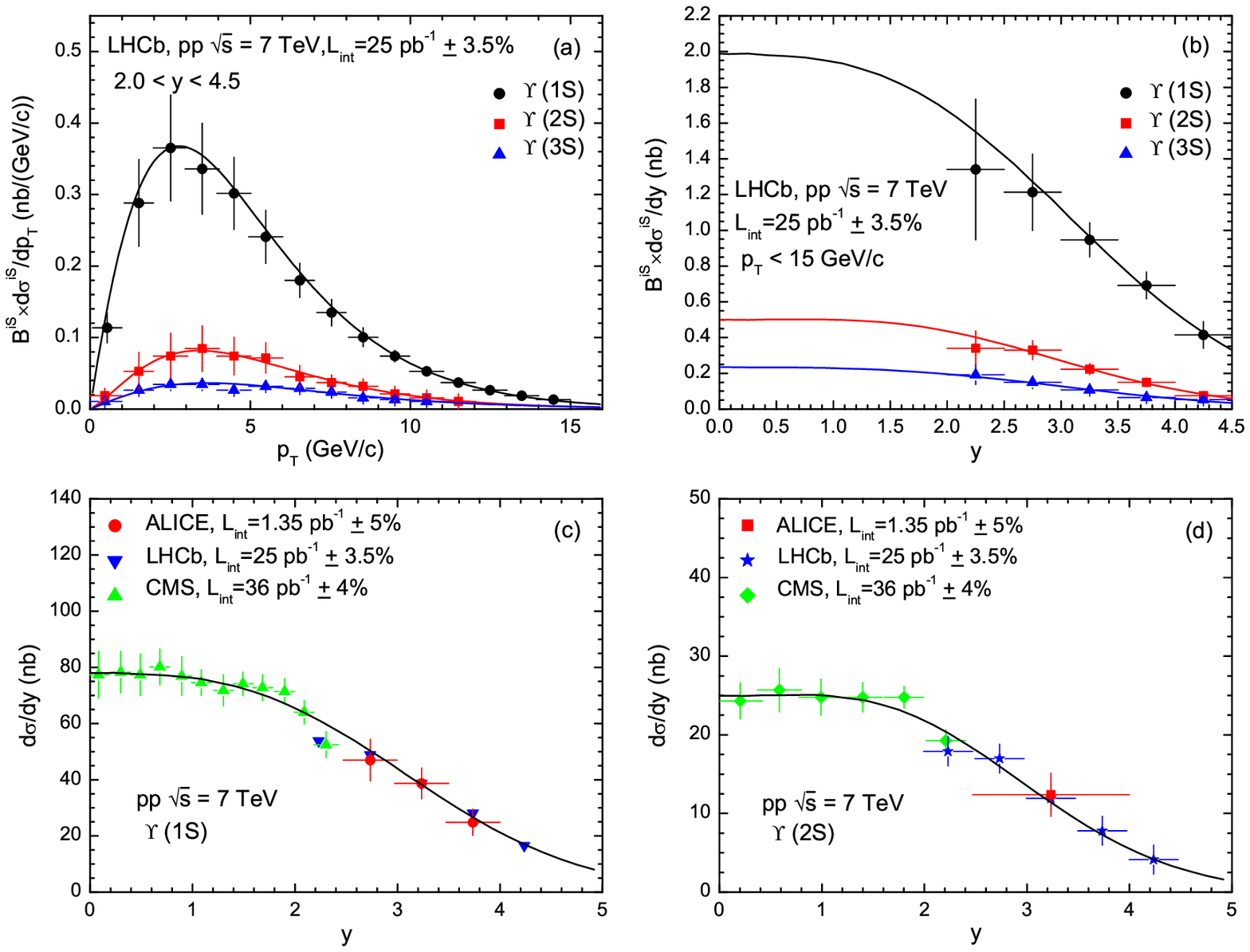}
\end{center}
\vskip-.20cm Fig. 8. (a) Transverse momentum spectra and (b)
backward (forward) region rapidity spectra of bottomonium $b\bar
b$ mesons $\Upsilon(1S)$, $\Upsilon(2S)$, and $\Upsilon(3S)$, as
well as whole space rapidity spectra of (c) $\Upsilon(1S)$ and (d)
$\Upsilon(2S)$ mesons produced in $pp$ collisions at $\sqrt{s}=7$
TeV, where $\sigma^{iS}$ and $B^{iS}$ on the vertical axis in
(a)(b) denote the cross-section and branching fractions of
$\Upsilon(iS)$ ($i=1$, 2, and 3), respectively. The closed symbols
represent the experimental data of the LHCb [35], ALICE [30], and
CMS Collaborations [36, 37]. The error bars in (a)(b) are the
total uncertainties which combine the systematic and statistical
uncertainties to be equal to their root-sum-square, and the error
bars in (c)(d) are only the systematic uncertainties. The curves
are our results calculated by using the hybrid model.
\end{figure}

\begin{figure}
\hskip-1.0cm \begin{center}
\includegraphics[width=10.0cm]{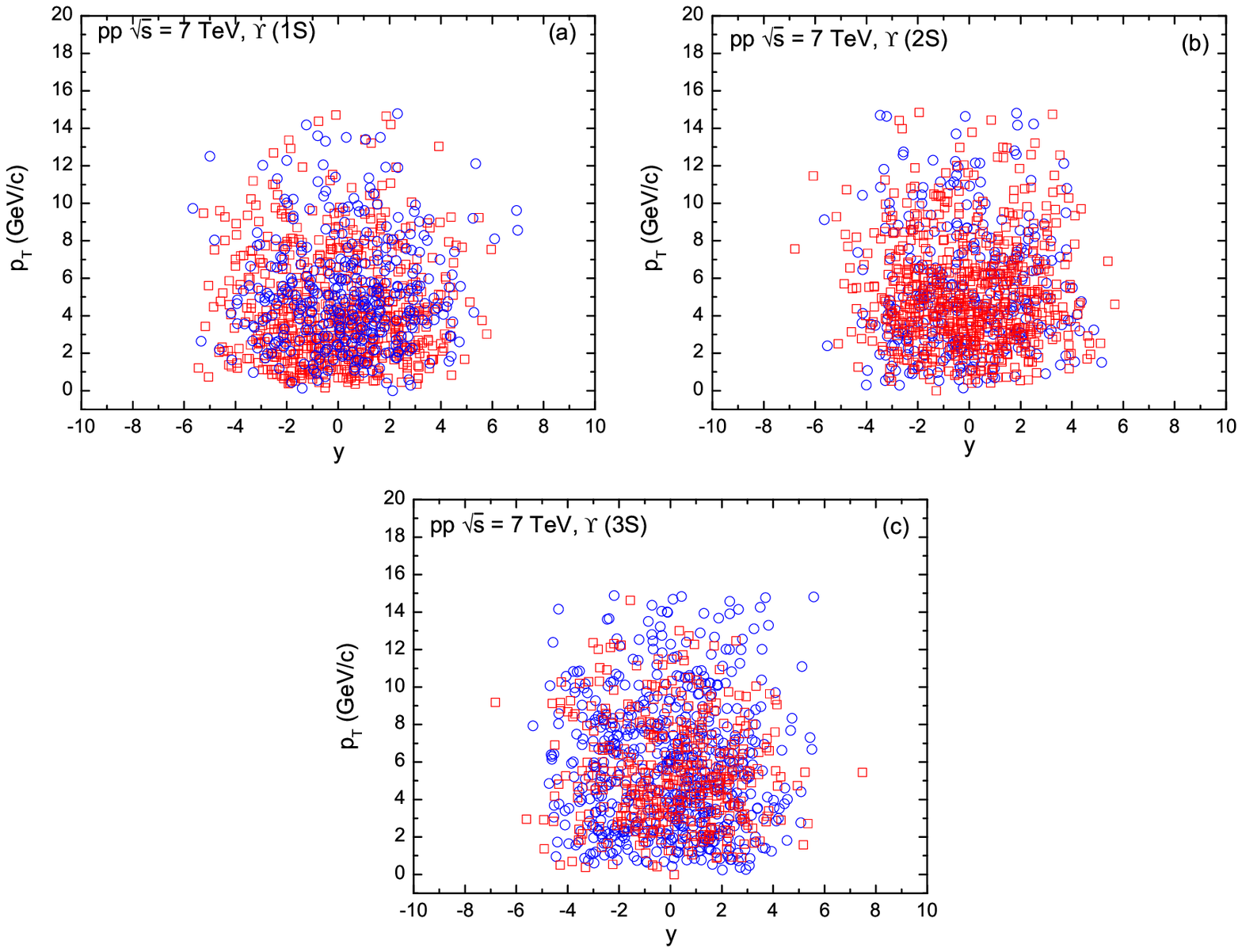}
\end{center}
\vskip-.20cm Fig. 9. Same as Figure 3, but showing the results for
(a) $\Upsilon(1S)$, (b) $\Upsilon(2S)$, and (c) $\Upsilon(3S)$
mesons produced in $pp$ collisions at $\sqrt{s}=7$ TeV.
\end{figure}

\begin{figure}
\hskip-1.0cm \begin{center}
\includegraphics[width=10.0cm]{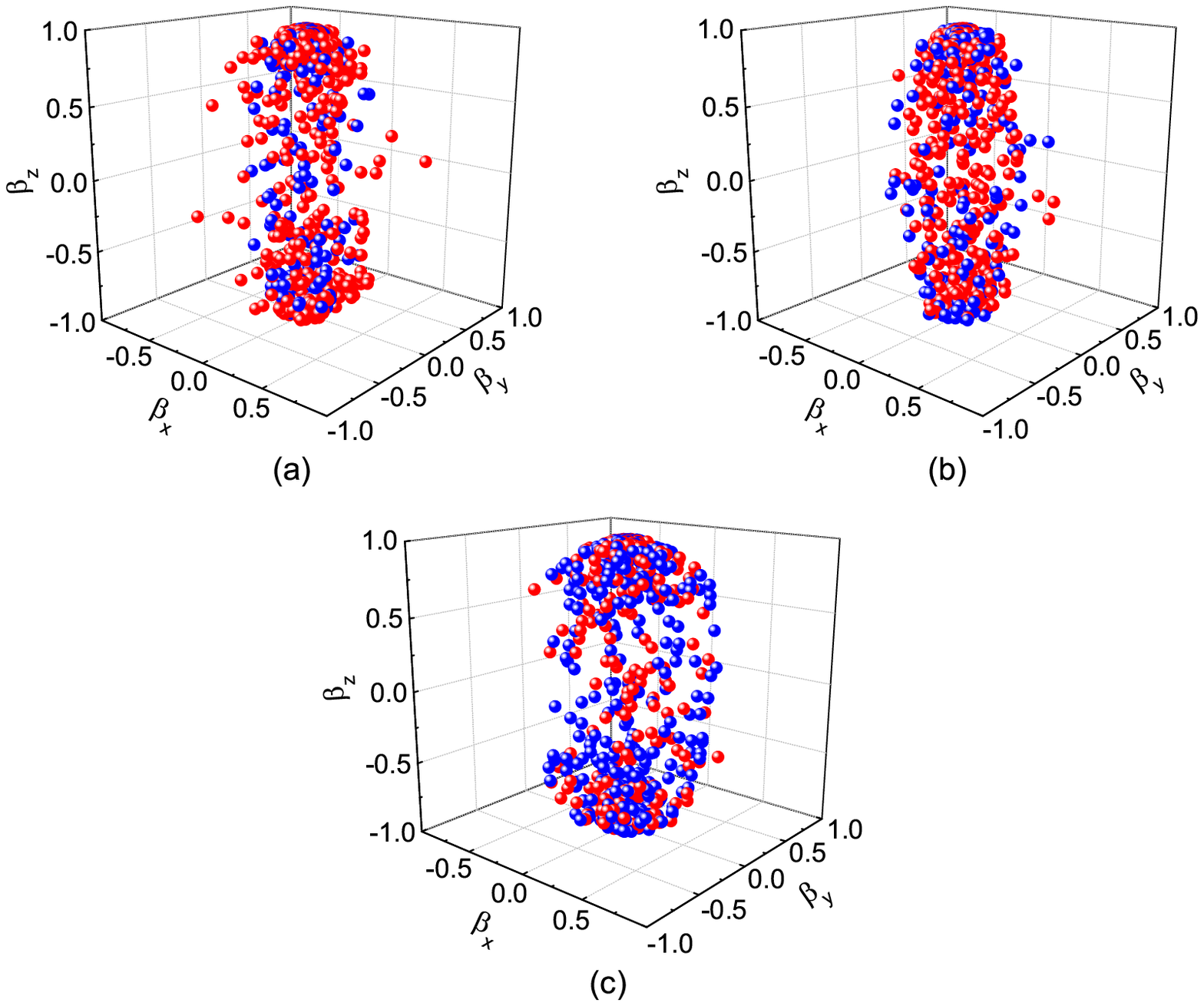}
\end{center}
\vskip-.20cm Fig. 10. Same as Figure 4, but showing the results
for (a) $\Upsilon(1S)$, (b) $\Upsilon(2S)$, and (c) $\Upsilon(3S)$
mesons produced in $pp$ collisions at $\sqrt{s}=7$ TeV.
\end{figure}

\begin{figure}
\hskip-1.0cm \begin{center}
\includegraphics[width=10.0cm]{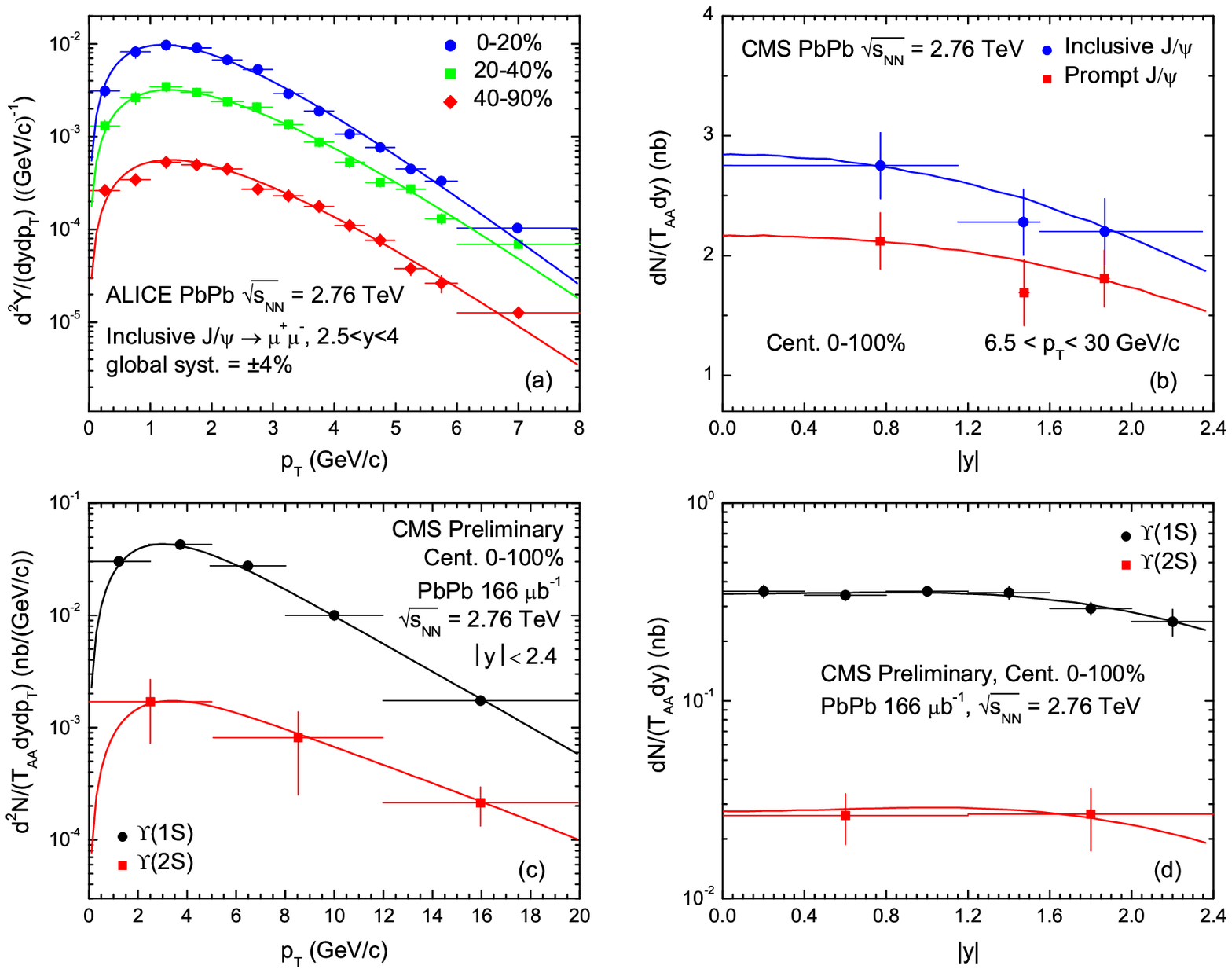}
\end{center}
\vskip-.20cm Fig. 11. (a) Transverse momentum spectra and (b)
rapidity spectra of $J/\psi$ mesons, as well as (c) transverse
momentum spectra and (d) rapidity spectra of $\Upsilon(1S)$ and
$\Upsilon(2S)$ mesons, produced in Pb-Pb collisions at
$\sqrt{s_{NN}}=2.76$ TeV, where $Y$ on the vertical axis denotes
the yield, and $T_{AA}$ denotes the nuclear overlap function which
depends on the collision centrality. The closed symbols represent
the (preliminary) experimental data of the ALICE [38] and CMS
Collaborations [39, 40] in different conditions shown in the
panels, and the error bars are only the systematic uncertainties.
The curves are our results calculated by using the hybrid model.
\end{figure}

The distributions of inclusive $J/\psi$ $p_T$, inclusive and
prompt $J/\psi$ $y$, $\Upsilon(1S)$ and $\Upsilon(2S)$ $p_T$, as
well as $\Upsilon(1S)$ and $\Upsilon(2S)$ $y$ in Pb-Pb collisions
at $\sqrt{s_{NN}}=2.76$ TeV are presented in Figures 11(a)--11(d),
respectively, where $Y$ on the vertical axis denotes the yield,
and $T_{AA}$ denotes the nuclear overlap function which depends on
the collision centrality. The closed symbols represent the
experimental data of the ALICE [38] and CMS Collaborations [39,
40] measured in different conditions shown in the panels, and the
error bars are only the systematic uncertainties. The curves are
our results calculated by using the hybrid model in which the
$p_T$ and $y$ spectra are described by the two-component Erlang
distribution and the two-component Gaussian distribution
respectively. The values of free parameters, normalization
constants, and $\chi^2$/dof are listed in Tables 1 and 2. One can
see that the results calculated by using the hybrid model are in
agreement with the experimental data of quarkonium states
($J/\psi$, $\Upsilon(1S)$, and $\Upsilon(2S)$) measured by the
ALICE and CMS Collaborations.

Because Figures 11(a) and 11(b) are the data set from different
experimental conditions, we do not extract the event patterns from
them. Based on the parameter values obtained from Figures 11(c)
and 11(d), we obtain some event patterns for (a) $\Upsilon(1S)$
and (b) $\Upsilon(2S)$ in Figures 12 and 13, where the event
patterns in $p_T-y$ and $\beta_x-\beta_y-\beta_z$ spaces are given
respectively. The red squares and globules represent the
contribution of the first component, and the blue circles and
globules represent the contribution of the second component, in
the two-component Erlang distribution. The corresponding values of
root-mean squares and the maximum velocity components are listed
in Tables 3 and 4. One can see the same or similar conclusion as
that from Figures 9 and 10. The maximum velocity surface is a fat
cylinder which has the length being 1.2--1.3 times of diameter.

We can give qualitatively a prediction on the prospects to
actually get $J/\psi$ rapidity distributions in Figure 11(b) for
the given centrality classes listed in Figure 11(a). From Table 1
we see that $\langle p_{Ti} \rangle_1$ and $\langle p_{Ti}
\rangle_2$ for Figure 11(a) increase slightly with decrease of the
centrality. As a non-leading particle, the rapidity distribution
of $J/\psi$ does not depend obviously on the centrality. These two
characteristics render that in the derived event pattern, the
particles will scatter slightly in higher $p_T$ region in
peripheral collisions. From central to peripheral collisions, the
increase in $p_T$ is slight due to the slight increases in
$\langle p_{Ti} \rangle_1$ and $\langle p_{Ti} \rangle_2$.

\begin{figure}
\hskip-1.0cm \begin{center}
\includegraphics[width=10.0cm]{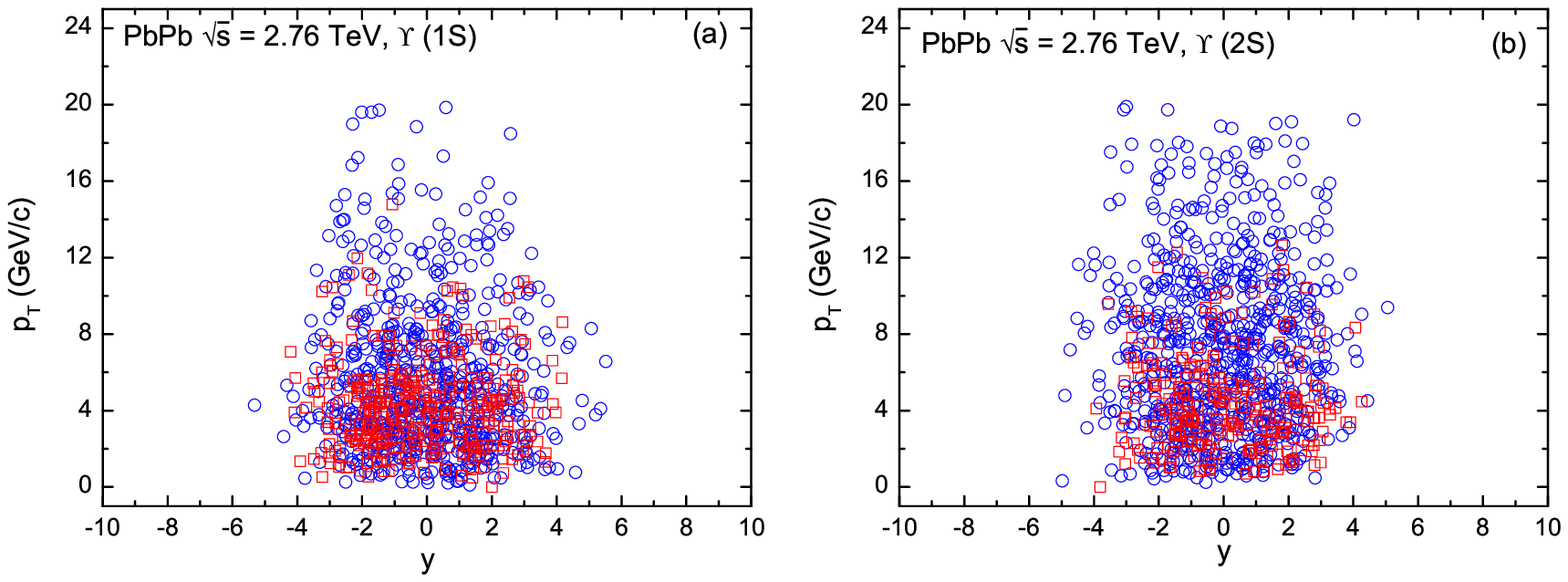}
\end{center}
\vskip-.20cm Fig. 12. Same as Figure 3, but showing the results
for (a) $\Upsilon(1S)$ and (b) $\Upsilon(2S)$ mesons produced in
Pb-Pb collisions at $\sqrt{s}=2.76$ TeV.
\end{figure}

\begin{figure}
\hskip-1.0cm \begin{center}
\includegraphics[width=10.0cm]{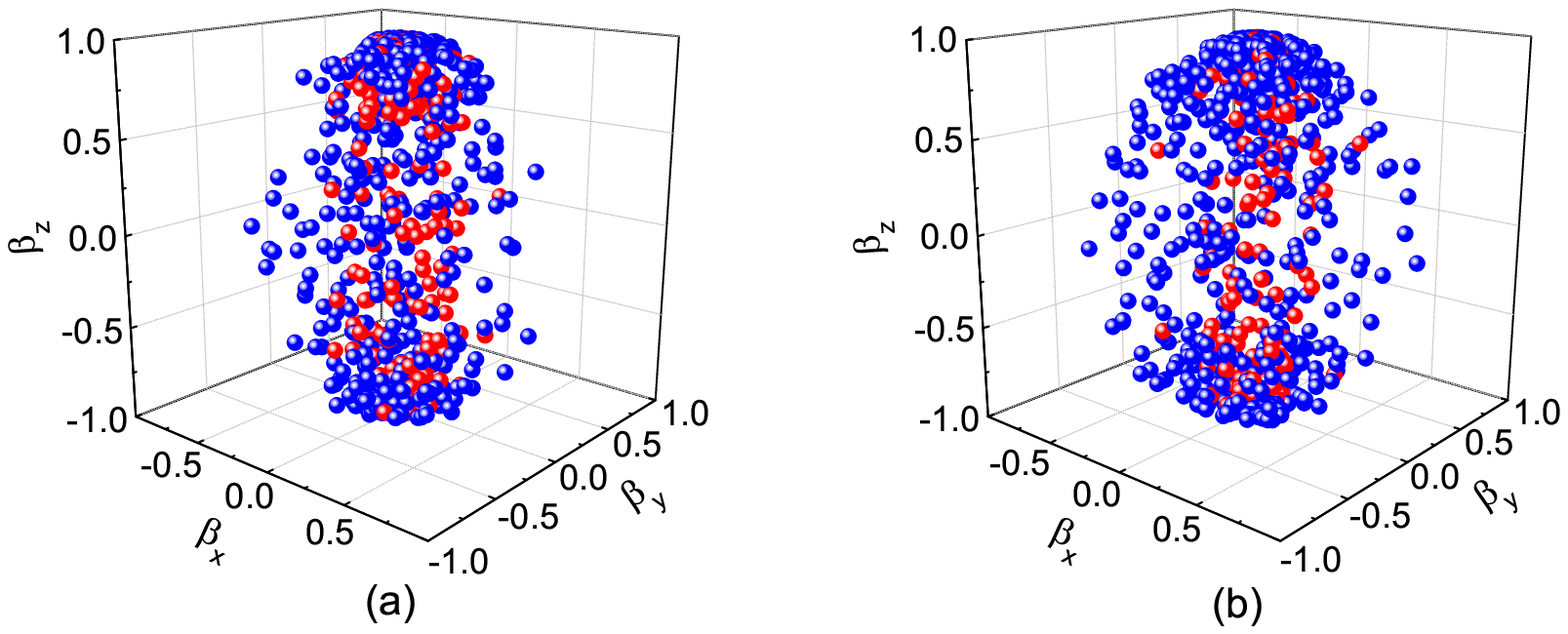}
\end{center}
\vskip-.20cm Fig. 13. Same as Figure 4, but showing the results
for (a) $\Upsilon(1S)$ and (b) $\Upsilon(2S)$ mesons produced in
Pb-Pb collisions at $\sqrt{s}=2.76$ TeV.
\end{figure}

The most important set of parameters in Table 1 are the mean
contributions, $\langle p_{Ti} \rangle_1$ and $\langle p_{Ti}
\rangle_2$, of each source for $p_T$ spectrum in the first and
second components. Then, the mean contribution of each source in
the first+second components is $\langle p_{Ti} \rangle = \langle
p_T \rangle /[k_1m_1+(1-k_1)m_2]$. The mean transverse momenta
contributed by the first and second components are $m_1 \langle
p_{Ti} \rangle_1$ and $m_2 \langle p_{Ti} \rangle_2$ respectively.
The most important set of parameters in Table 2 are the peak
position $y_F$ ($y_B$) and distribution width $\sigma_{yF}$
($\sigma_{yB}$) for the forward (backward) region in the
two-component Gaussian function for $y$ spectrum. To see the
dependences of these parameters on $m_0$, Figure 14 shows the
relations between (a) $\langle p_{Ti} \rangle$ ($\langle p_{Ti}
\rangle_1$, $\langle p_{Ti} \rangle_2$) and $m_0$, (b) $\langle
p_T \rangle$ ($m_1 \langle p_{Ti} \rangle_1$, $m_2 \langle p_{Ti}
\rangle_2$) and $m_0$, (c) $y_F$ and $m_0$, as well as (d)
$\sigma_{yF}$ and $m_0$, in $pp$ and Pb-Pb collisions. Different
symbols represent different parameters for different collisions
shown in the panels. One can see an obvious increase in $\langle
p_{Ti} \rangle$, $\langle p_{Ti} \rangle_1$, $\langle p_{Ti}
\rangle_2$, $\langle p_T \rangle$, $m_1 \langle p_{Ti} \rangle_1$,
and $m_2 \langle p_{Ti} \rangle_2$, and a decrease in $y_F$ and
$\sigma_{yF}$, when $m_0$ changes from $\sim$3 GeV to $\sim$90
GeV. This mass scale imposes a difference on the observed data
from the presence of a quantum chromodynamics (QCD) hard scale.

\begin{figure}
\hskip-1.0cm \begin{center}
\includegraphics[width=10.0cm]{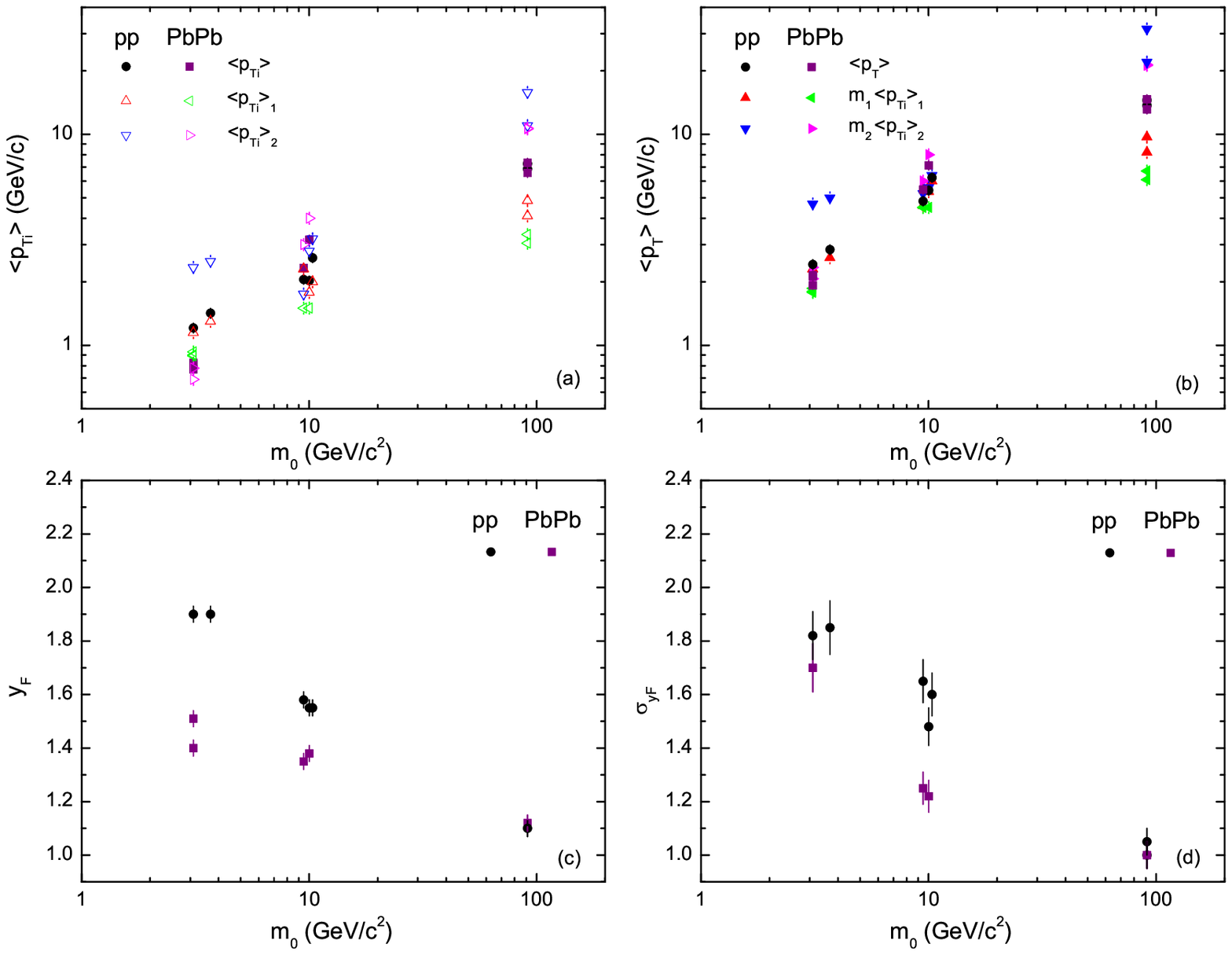}
\end{center}
\vskip-.20cm Fig. 14. Relations between (a) $\langle p_{Ti}
\rangle$ ($\langle p_{Ti} \rangle_1$, $\langle p_{Ti} \rangle_2$)
and $m_0$, (b) $\langle p_T \rangle$ ($m_1 \langle p_{Ti}
\rangle_1$, $m_2 \langle p_{Ti} \rangle_2$) and $m_0$, (c) $y_F$
and $m_0$, as well as (d) $\sigma_{yF}$ and $m_0$, in $pp$ and
Pb-Pb collisions. Different symbols represent different parameters
for different collisions shown in the panels, where $\langle
p_{Ti} \rangle$, $\langle p_{Ti} \rangle_1$, and $\langle p_{Ti}
\rangle_2$ denote the source's mean contributions in the
first+second, first, and second components, respectively; and
$\langle p_T \rangle$, $m_1 \langle p_{Ti} \rangle_1$, and $m_2
\langle p_{Ti} \rangle_2$ denote the mean transverse momenta
contributed by the first+second, first, and second components,
respectively.
\end{figure}

From Table 1 one can see that in some cases $k_1$ is less than
0.5, and $\langle p_{Ti} \rangle_1$ is greater than $\langle
p_{Ti} \rangle_2$. These results does not mean a reversal of the
first and second components, or a reversal of the low- and
high-$p_T$ regions. In fact, the first component that contributes
in the low-$p_T$ region is determined by both $m_1$ and $\langle
p_{Ti} \rangle_1$, and the second component that contributes in
the high-$p_T$ region is determined by both $m_2$ and $\langle
p_{Ti} \rangle_2$. The mean transverse momenta, $m_1 \langle
p_{Ti} \rangle_1$ and $m_2 \langle p_{Ti} \rangle_2$, contributed
by the first and second components, respectively, play a
determinate action. Generally, we restrict $m_1 \langle p_{Ti}
\rangle_1 < m_2 \langle p_{Ti} \rangle_2$, which ensures the first
component contributing in the low-$p_T$ region and the second
component contributing in the high-$p_T$ region. The size of $k_1$
is not related to the relative sizes of $m_1 \langle p_{Ti}
\rangle_1$ and $m_2 \langle p_{Ti} \rangle_2$.

We would like to point out that $Z$ bosons and quarkonium states
are chosen by us in the present work due to their productions
being through hard scattering and at initial state, which are
different from those of light flavor particles which are mostly
produced through soft excitation and at intermediate state.
Although we describe uniformly the presence of a variable mass
scale from 3 to 90 GeV, we cannot provide further information in
theory. In fact, the method used in the present work is
independent of models, which is only based on the description of
probability distribution of experimental data. If we do not use
formulas, but discrete probabilities based on data, similar event
patters can be obtained. In addition, our choices on data sets are
not sole. In fact, other data sets which contained both $p_T$ and
$y$ spectra can be used to structure the event patterns.

It should be noted that in the calculation of $\chi^2$ in Figures
1, 2, 5 ,8 and 11, we have to determine the corresponding relation
between a theoretical point and an experimental data in a given
bin of a variable. According to ref. [41], there are more than one
methods to determine the corresponding relation, which include
considerations in variable and frequency respectively. In the
present work, we have simply used the method of histograms, i.e.
presenting the theoretical point and experimental data in the form
of histograms when we calculate $\chi^2$. In most cases, if the
bin is narrow enough, to use the closest theoretical point and
experimental data is an approximate alternative method.

It should also be noted that Figures 3, 4, 6, 7, 9, 10, 12, 13 are
supposed to only give pictorial representations of the particle
scatter plots at the stage of kinetic freeze-out of the
interacting system. These pictorial representations are based on
the experimental data themselves. They are independent of models
due to Eqs. (3) and (5) being only parameterizations of the
experimental data, and Eqs. (7), (8), (10), and (11) being only
stochastic extractions based on Eqs. (3) and (5). Any other event
generators that describe simultaneously the spectra of $p_T$ and
$y$ can give similar pictorial representations. As we know,
concrete and ready-made pictorial representations of the particle
scatter plots at the stage of kinetic freeze-out given by other
event generators are not available at present.

It seems that possible effects from kinematic cuts are not
considered for the extracted discrete values, then for the event
patterns. As we know, all the experimental data are subject to
trigger and event selection restrictions, which have subsequent
effects on the kinematic region in which the cross sections are
defined. In fact, because our analyses are based on the
experimental data themselves, the extracted parameters imply these
restrictions. This renders that there are considerations on the
impacts of these restrictions on the subsequent event shapes.
Because the differences in experimental restrictions for $pp$ and
Pb-Pb collisions as well as for spectra of $Z$ bosons and
quarkonium states are small, the impacts on the values of
parameters, discrete values, and event patterns are small in some
details. Generally, these small differences in restrictions do not
effect largely the global shapes and contour profiles of events.
\\

{\section{Conclusions}}

We summarize here our main observations and conclusions.

(a) The transverse momentum and rapidity spectra of $Z$ bosons
produced in $pp$ and Pb-Pb collisions at 2.76 TeV, the same
spectra of charmonium $c\bar c$ mesons ($J/\psi$ and $\psi(2S)$)
and bottomonium $b\bar b$ mesons ($\Upsilon(1S)$, $\Upsilon(2S)$,
and $\Upsilon(3S)$) produced in $pp$ collisions at 7 TeV, and the
same spectra of $J/\psi$, $\Upsilon(1S)$, and $\Upsilon(2S)$
produced in Pb-Pb collisions at 2.76 TeV at the LHC are uniformly
described by the hybrid model of two-component Erlang distribution
for $p_T$ spectrum and the two-component Gaussian distribution for
$y$ spectrum. The former distribution results from the multisource
thermal model, and the later one results from the revised Landau
hydrodynamic model. The modelling results are in agreement with
the experimental data measured at the LHC.

(b) In the two-component Erlang distribution Eq. (3), the first
component that contributes in the low-$p_T$ region is determined
by both $m_1$ and $\langle p_{Ti} \rangle_1$, and the second
component that contributes in the high-$p_T$ region is determined
by both $m_2$ and $\langle p_{Ti} \rangle_2$. The mean transverse
momenta contributed by the first and second components are $m_1
\langle p_{Ti} \rangle_1$ and $m_2 \langle p_{Ti} \rangle_2$
respectively, and the mean transverse momentum contributed by the
two components is $\langle p_T \rangle = k_1 m_1 \langle p_{Ti}
\rangle_1 + (1-k_1)m_2 \langle p_{Ti} \rangle_2$. Generally, we
restrict $m_1 \langle p_{Ti} \rangle_1 < m_2 \langle p_{Ti}
\rangle_2$, which ensures the first component contributing in the
low-$p_T$ region and the second component contributing in the
high-$p_T$ region. The two-component Gaussian function Eq. (5) is
a revision of the Landau hydrodynamic model.

(c) The most important set of parameters are the mean
contributions, $\langle p_{Ti} \rangle_1$ and $\langle p_{Ti}
\rangle_2$, of each source in the first and second components of
the two-component Erlang distribution for $p_T$ spectrum, and the
peak position $y_F$ ($y_B$) and distribution width $\sigma_{yF}$
($\sigma_{yB}$) for the forward (backward) region in the
two-component Gaussian function for $y$ spectrum. The mean
contribution of each source in the two-component Erlang
distribution is $\langle p_{Ti} \rangle = \langle p_T \rangle
/[k_1m_1+(1-k_1)m_2]$. From the dependences of related parameters
on $m_0$, one can see an obvious increase in $\langle p_{Ti}
\rangle$, $\langle p_{Ti} \rangle_1$, $\langle p_{Ti} \rangle_2$,
$\langle p_T \rangle$, $m_1 \langle p_{Ti} \rangle_1$, and $m_2
\langle p_{Ti} \rangle_2$, and a decrease in $y_F$ and
$\sigma_{yF}$, when $m_0$ changes from $\sim$3 GeV to $\sim$90
GeV. This mass scale imposes a difference on the observed data
from the presence of a QCD hard scale. Because of some non-violent
collision nucleons existing in the overlapping region, Pb-Pb
collisions show more or less lower parameter values than $pp$
collisions in some cases.

(d) Based on the parameter values extracted from $p_T$ and $y$
spectra, the event patterns (or particle scatter plots) in
two-dimensional $p_T$-$y$ space and in three-dimensional velocity
(or coordinate) space are obtained. As results of hard scattering
process at the initial state, both the $Z$ bosons and quarkonium
states present some similarities and differences in event
patterns. Generally, $\sqrt{\overline{\beta_x^2}} \approx
\sqrt{\overline{\beta_y^2}} \ll \sqrt{\overline{\beta_z^2}}$, and
$|\beta_x|_{\max} \approx |\beta_y|_{\max}<|\beta_z|_{\max}$,
which renders that the event pattern in velocity (or coordinate)
space looks like a rough cylinder along the beam direction when it
emits $Z$ bosons and quarkonium states, and the maximum velocity
surface is a fat cylinder which has the length being 1.2--2.2
times of diameter. The situations for $pp$ and Pb-Pb collisions do
not show an obvious difference.

(e) The cylinder diameters ($2\sqrt{\overline{\beta_x^2}}$ or
$2|\beta_x|_{\max}$) or lengths ($2\sqrt{\overline{\beta_z^2}}$ or
$2|\beta_z|_{\max}$) corresponding to emissions of charmonium
$c\bar c$ mesons and bottomonium $b\bar b$ mesons are nearly the
same. Because of large mass, $Z$ bosons correspond to a thinner
and shorter cylinder than quarkonium states in terms of
$2\sqrt{\overline{\beta_x^2}}$ and $2\sqrt{\overline{\beta_z^2}}$
in velocity space, though the values of $|\beta_z|_{\max}$ for
both the types of particles are almost the same. Meanwhile, the
values of $\sqrt{\overline{y^2}}$ for $Z$ bosons are less than
those for quarkonium states due to the shorter cylinder
corresponding to $Z$ bosons. The values of
$\sqrt{\overline{p_T^2}}$ for $Z$ bosons are greater than those
for quarkonium states due to the excitation degree of emission
sources for $Z$ bosons being higher than that for quarkonium
states. The situations for $pp$ and Pb-Pb collisions do not show
an obvious difference.

(f) The present work provides a reference in methodology which
displays event patterns (particle scatter plots) for different
particles in three-dimensional velocity space or other available
spaces based on the transverse momentum and rapidity spectra of
considered particles. Because the present analyses are based on
the experimental data themselves, the extracted parameters imply
experimental restrictions. This renders that the discrete values
and event patterns extracted by us are accurate to the best of our
ability. It is expected that different particles correspond to
different event patterns in different shapes with different sizes.
To give a better comparison, the same experimental restrictions
for different particles are needed in the future.
\\

{\bf Conflict of Interests}

The authors declare that there is no conflict of interests
regarding the publication of this paper.
\\

{\bf Acknowledgments}

This work was supported by the National Natural Science Foundation
of China under Grant No. 11575103 and the US DOE under contract
DE-FG02-87ER40331.A008.

\end{document}